\documentclass[
 aip,
 amsmath,amssymb,
 reprint,%
]{revtex4-1}

\usepackage{graphicx}
\usepackage{dcolumn}
\usepackage{bm}
\newcommand{\nodata}{\textellipsis}
\usepackage[utf8]{inputenc}
\usepackage[T1]{fontenc}
\usepackage{mathptmx}
\usepackage{amsmath}
\usepackage{etoolbox}
\usepackage{lineno}

\usepackage{longtable}
\usepackage{booktabs}
\usepackage{threeparttable}
\usepackage{footnote}

\makeatletter
\def\@email#1#2{%
 \endgroup
 \patchcmd{\titleblock@produce}
  {\frontmatter@RRAPformat}
  {\frontmatter@RRAPformat{\produce@RRAP{*#1\href{mailto:#2}{#2}}}\frontmatter@RRAPformat}
  {}{}
}%
\makeatother
\begin{document}

\preprint{FDU-IMP-2025-01}

\title{Atomic data benchmarked by Large-scale Multiconfiguration Dirac-Hartree-Fock Calculations for Beryllium}
\author{S. J. Wu}
\affiliation{Shanghai EBIT Lab, Key Laboratory of Nuclear Physics and Ion-beam Application, Institute of Modern Physics, Department of Nuclear Science and Technology, Fudan University, Shanghai 200433, China}
\author{S. W. Tian}
\affiliation{Shanghai EBIT Lab, Key Laboratory of Nuclear Physics and Ion-beam Application, Institute of Modern Physics, Department of Nuclear Science and Technology, Fudan University, Shanghai 200433, China}
\author{R. Si}
\email{rsi@fudan.edu.cn}
\affiliation{Shanghai EBIT Lab, Key Laboratory of Nuclear Physics and Ion-beam Application, Institute of Modern Physics, Department of Nuclear Science and Technology, Fudan University, Shanghai 200433, China}
\author{K. Wang}
\affiliation{Department of Physics and Anhui Key Laboratory of Optoelectric Materials Science and Technology, Key Laboratory of Functional Molecular Solids, Ministry of Education, Anhui Normal University, Wuhu, Anhui 241000, China}
\author{P. J{\"o}nsson}
\affiliation{Department of Materials Science and Applied Mathematics, Malm{\"o} University, SE-20506, Malm{\"o}, Sweden}
\author{G. Gaigalas}
\affiliation{Institute of Theoretical Physics and Astronomy, Vilnius University, Saul$\dot{e}$tekio Avenue 3, LT-10222, Vilnius, Lithuania}
\author{M. Godefroid}
\affiliation{Spectroscopy, Quantum Chemistry and Atmospheric Remote Sensing, CP160/09, Universit\'e libre de Bruxelles, B-1050 Brussels, Belgium}
\author{A. M. Amarsi}
\affiliation{Theoretical Astrophysics, Department of Physics and Astronomy, Uppsala University, Box 516, SE-751 20 Uppsala, Sweden}
\author{C. Y. Chen}
\email{chychen@fudan.edu.cn}
\affiliation{Shanghai EBIT Lab, Key Laboratory of Nuclear Physics and Ion-beam Application, Institute of Modern Physics, Department of Nuclear Science and Technology, Fudan University, Shanghai 200433, China}

\date{\today}

\begin{abstract}
 The multiconfiguration Dirac-Hartree-Fock (MCDHF) and relativistic configuration interaction (RCI) methods are used to provide excitation energies, radiative transition data, lifetimes, Land$\acute{\rm{e}}$ $g$-factors, hyperfine interaction constants and isotope shift parameters for the 99 lowest levels of configurations $1s^2 2s nl$ ($n \leq 7$) + $1s^2 2p^2$ in beryllium. Compared with available experimental excitation energies, the average difference with the standard deviation is 7.08 $\pm$ 1.14 $\rm{cm^{-1}}$ (0.011$\%$ $\pm$ 0.003$\%$), which demonstrates the excellent theory-observation agreement. The uncertainties of the transition rates are estimated based on two independent methods. The present MCDHF/RCI oscillator strengths and those obtained from the explicitly correlated Gaussian (ECG) method all agree within 2\%, except for four transitions affected by strong cancellation effects. For lifetimes, hyperfine splittings and isotope shifts, the present MCDHF/RCI results show good agreement with the few available experimental values, supporting the reliability of our predictions for many states lacking experimental measurements. These comprehensive results can be used in line identification and diagnostics of astrophysical plasmas.
 
\end{abstract}

\maketitle

\section{Introduction} \label{sec:intro}

Atomic spectroscopic data are essential to many astrophysical studies, where the lighter elements with few electrons have attracted much attention~\citep{Kallman.2007.p79,DelZanna.2018.p}. Through analyzing atomic spectra recorded from astrophysical observations, properties such as abundances of elements, electron temperature, electron density, along with important excitation mechanisms and photoprocesses can be inferred~\citep{Traebert.2014.p6,Traebert.2014.p14}. Beryllium has long been a topic of interest for astrophysical observations~\citep{Traebert.2014.p114003,Buldgen.2023.p9,Boesgaard.2023.p40} and its spectrum has been extensively studied. Comprehensive lists of previously published research on Be I energy levels (more than 100 papers) and transition probabilities (more than 200 papers) can be found in the bibliographic databases of the National Institute of Standards and Technology (NIST)~\citep{Kramida.2023.p}. Beryllium abundances in late-type stars serve as a crucial diagnostic in a variety of astrophysical contexts~\citep{Molaro.2020.p2902,Boesgaard.2022.p21} and can be used in studies of evolutionary mixing, galactic chemical evolution, planet engulfment, and the formation of globular clusters~\citep{Smiljanic.2010.p50,TucciMaia.2015.p10,Desidera.2016.p46,Prantzos.2012.p67,Giribaldi.2022.p117}.

In order to meet the requirements for accurate analysis of astrophysical observations, high-resolution atomic spectra obtained from laboratory measurements and theoretical calculations are always required. Unfortunately, relevant data provided so far have often been insufficiently complete or not accurate enough for the analysis. Considerable research has been carried out in these areas for many years.

Relevant experimental measurements of the beryllium spectrum date back to the end of the 19th century. The wavelength measurements, which extend from 89 to 31780 \AA, were obtained from emission spectra of hollow-cathode and arc discharges, beam-foil spectra and far-ultraviolet photo-absorption spectra. For a review of the research before 1997, details can be found in the compilation done by~\citet{Kramida.1997.p1185}, where the work of Johansson and Holmstr{\"o}m~\citep{JOHANSSON.1962.p119,Johansson.1974.p236,Holmstroem.1969.p133} provided many accurate results. Most of the energy levels compiled in NIST Atomic Spectra Database (ASD)~\citep{Kramida.2023.p} are from these three works. Apart from these studies,~\citet{Cook.2018.p} have recently reported spectroscopic results on the $2s2p~^{1}P_1$ state in Be I through doppler-free spectroscopy obtained from an ultra-low expansion optical cavity. The absolute frequency for the center of gravity is determined to be $42~565.4501(13)$ $\rm{cm}^{-1}$, a factor of 130 more precise than the previous experimental measurement. Two years later, an absolute frequency measurement of the $2s3d~^{1}D_2$ state in neutral $^{9}\rm{Be}$, which is a factor of 180 more precise than the previous experimental measurement, was also made by Cook $et~al$.~\cite{Cook.2020.p} using two-photon spectroscopy with a resonant intermediate state. The absolute energy of the $2s3s~^{1}S_0$ state in neutral $^{9}\rm{Be}$ was recently measured by Ahrendsen $et~al.$~\cite{Ahrendsen.2024.p}, who improved on the precision of the previous experimental result by a factor of 20. All these precise experimental measurements have, to some extent, enriched the reference data for neutral beryllium.

A significant body of work exists on theoretical calculations of the low-lying states in Be I, as evidenced by the extensive listings in the bibliographic databases associated with NIST ASD~\citep{Kramida.2023.p}. The following review does not attempt to be exhaustive but rather focuses on discussing a selection of key studies that are most relevant to the context and advancements discussed in the present work. Excitation energies, lifetimes, and transition probabilities for $n\leq3$ levels were calculated for the Be sequence using the multiconfiguration Hartree-Fock (MCHF) method with relativistic effects included through the Breit-Pauli (BP) Hamiltonian by Froese Fischer and Tachiev~\citep{Tachiev.1999.p5805, FroeseFischer.2004.p1}. Through the Hylleraas-configuration interaction (Hy-CI) method, Sims et al.~\citep{Sims.2011.p,Sims.2014.p,Sims.2020.p} calculated nonrelativistic energies for the $2~^1S$ ground state and $3~^1S$, $4~^1S$, $5~^1S$, $6~^1S$, $7~^1S$ excited states of the beryllium atom.~\citet{Dong.2021.p43103} calculated excitation energies, transition probabilities and polarizabilities for a few low-lying states of Be I through the fully relativistic multiconfiguration Dirac-Hartree-Fock (MCDHF) method. In recent years, using variationally optimized Gaussian functions that explicitly depend on the distances between the electrons, i.e., the so-called explicitly correlated Gaussian functions (ECGs), the transition energies of low-lying states for Be I were calculated with high accuracy by Stanke $et ~al.$~\citep{Stanke.2007.p,Stanke.2007.pa,Stanke.2009.p,Stanke.2018.p,Stanke.2019.p,Stanke.2019.p155002,Stanke.2021.p138823,Stanke.2022.p,Kedziorski.2020.p137476} and Nasiri $et~al.$~\cite{Nasiri.2021.p}. Their work mainly involved the $^1S$, $^1P^{\circ}$, $^1D$ and $^3P^{\circ}$ terms of Be I. The deviation between their results and experimental values is generally less than 0.1 $\rm{cm^{-1}}$, which is highly satisfactory. 
However, it usually requires several months or even one year of continuous computing with hundreds of cores on a parallel computer system for one or two terms. Therefore, the number of energy levels involved in their calculations is usually not large. So far, they have only provided transition rates between the ground and 19 lowest bound excited singlet $S$ and $P^{\circ}$ states of the beryllium atom~\citep{Nasiri.2021.p}.

As a part of our series of work on accurate atomic data~\citep{Wang.2018.p27,Wang.2018.p30,Wang.2018.p40,Wang.2018.p127,Wang.2019.p1,Song.2020.p70,Zhang.2021.p56,Li.2022.p50}, we performed large-scale MCDHF and relativistic configuration interaction (RCI) calculations using the {\sc Graspg} package~\citep{Si.2025.p109604}, which is a further optimized and improved version of the {\sc Grasp} code~\citep{Joensson.2013.p2197, FroeseFischer.2019.p184}. Relevant improvements are detailed in~\citet{Li.2023.p12,Li.2023.p108562} and~\citet{Si.2025.p109604}. The present study represents the first application of the improved program to the neutral beryllium atom. By demonstrating the accuracy and reliability of this approach, we aim to establish a robust framework for investigating atomic parameters of other neutral and lowly-charged atoms of astrophysical interest. Future work will extend this methodology to a series of atoms, starting from boron, and provide a comprehensive and systematic exploration of their atomic structures and properties. These results will be presented in a series of publications, contributing to high-precision studies in laboratory and astrophysical plasmas.

The calculations of this work provide highly accurate data of excitation energies, wavelengths, transition rates (electric dipole (E1) and quadrupole (E2), magnetic dipole (M1) and quadrupole (M2)), lifetimes, Land$\acute{\rm{e}}$ $g$-factors, hyperfine interaction constants and isotope shifts
for the 99 lowest levels of configurations $1s^2 2s nl$ ($n \leq 7$) + $1s^2 2p^2$
in \mbox{Be I}. This work is a continuation of our previous calculations~\citep{Wang.2018.p40}, in which a complete and accurate data set of excitation energies, lifetimes, and transition rates for the $ n\leq$ 6 levels in Be-like ions from B II to Ne VII are reported. To assess the accuracy of the present MCDHF and RCI data, two uncertainty estimation methods are used and extensive comparisons are made with available experimental and theoretical values for neutral beryllium. 

\section{Theory and Calculations} \label{sec:style}

\subsection{Atomic state functions} \label{subsec:mcdhf}
 
In the MCDHF approach, the atomic state functions (ASFs) are expressed as linear expansions of configuration state functions (CSFs). CSFs are built from antisymmetrized and coupled products of one-electron Dirac orbitals. This method was described by~\citet{Fischer.2016.p182004} in detail. Therefore, only computational procedures are described here.

In our MCDHF calculations, the 99 lowest levels of configurations $1s^2 2s nl$ ($n \leq 7$) + $1s^2 2p^2$ are targeted. The original multireference (MR) sets for even and odd parities include the following configurations
\begin{itemize}
    \item [even
    :]
    $2s^2$, $2p^2$, $2s 3s$, $2s 3d$, $2p 3p$, $2s 4s$, $2s 4d$, $2p 4p$, $2p 4f$, $2s 5s$, $2s 5d$, $2s 5g$, $2p 5p$, $2p 5f$, $2s 6s$, $2s 6d$, $2s 6g$, $2p 6p$, $2p 6f$, $2p 6h$, $2s 7s$, $2s 7d$, $2s 7g$, $2s 7i$, $2p 7p$, $2p 7f$, $2p 7h$, $2s 8s$, $2p 8p$; 
    \item [odd
    :]
    $2s 2p$, $2s 3p$, $2p 3s$, $2p 3d$, $2s 4p$, $2s 4f$, $2p 4s$, $2p 4d$, $2s 5p$, $2s 5f$, $2p 5s$, $2p 5d$, $2p 5g$, $2s 6p$, $2s 6f$, $2s 6h$, $2p 6p$, $2p 6f$, $2p 6g$, $2s 7p$, $2s 7f$, $2s 7h$, $2p 7s$, $2p 7d$, $2p 7g$, $2p 7i$, $2s 8p$, $2p 8s$.
\end{itemize}
These MR sets are subsequently referred to as $\rm{MR_0}$. Initial MCDHF calculations for the MR sets are performed simultaneously for even and odd parities to determine a common set of orbitals from $1s$ to $8p$, where two orbitals, $8s$ and $8p$, are introduced specifically to account for the $LS$-term dependence and optimized together with $1s,2s,2p,...,7i$ spectroscopic orbitals~\citep{FroeseFischer.1998.p1753,Papoulia.2019.p106}. 
An active space (AS) approach~\citep{Olsen.1988.p2185,Sturesson.2007.p539} is used to generate CSF expansions that are obtained by allowing single and double excitations from the originally occupied valence subshells to additional excited orbitals. The $1s$ subshell is kept closed in all CSFs of the expansions.
The first orbital active set considered is the following set: \\
$\rm{AS_1}$ = $\{8s,8p,8d,8f,8g,8h,8i,8k\}$. \\
Additional ASs are introduced through the following sequence:

$\rm{AS_2}$ = $\rm{AS_1}$ + $\{9s,9p,9d,9f,9g,9h,9i,9k,9l\}$,

$\rm{AS_3}$ = $\rm{AS_2}$ + $\{10s,10p,10d,10f,10g,10h,10i,10k,10l,10m\}$,

$\rm{AS_4}$ = $\rm{AS_3}$ + $\{11s,11p,11d,11f,11g,11h,11i,11k,11l,11m\}$,

$\rm{AS_5}$ = $\rm{AS_4}$ + $\{12s,12p,12d,12f,12g,12h,12i,12k,12l\}$,

$\rm{AS_6}$ = $\rm{AS_5}$ + $\{13s,13p,13d,13f,13g,13h,13i,13k,13l\}$,

$\rm{AS_7}$ = $\rm{AS_6}$ + $\{14s,14p,14d,14f,14g,14h,14i\}$,

$\rm{AS_8}$ = $\rm{AS_7}$ + $\{15s,15p,15d,15f,15g,15h,15i\}$,

$\rm{AS_9}$ = $\rm{AS_8}$ + $\{16s,16p,16d,16f,16g\}$,

$\rm{AS_{10}}$ = $\rm{AS_9}$ + $\{17s,17p,17d,17f\}$,

$\rm{AS_{11}}$ = $\rm{AS_{10}}$ + $\{18s,18p,18d\}$,

$\rm{AS_{12}}$ = $\rm{AS_{11}}$ + $\{19s,19p,19d\}$,

$\rm{AS_{13}}$ = $\rm{AS_{12}}$ + $\{20s,20p\}$,

$\rm{AS_{14}}$ = $\rm{AS_{13}}$ + $\{21s,21p\}$,

$\rm{AS_{15}}$ = $\rm{AS_{14}}$ + $\{22s,22p\}$.

The final AS is restricted to $\rm{AS_{15}}$ due to the fact that the number of orbitals in the {\sc Graspg} package is not allowed to exceed 214. This is the maximum number of orbitals that are possible with the packed labeling method used in {\sc Graspg}, in which the labels are 4-byte integers constructed from several separately encoded parts~\citep{Li.2023.p12}. The AS is systematically enlarged according to a sequence of extensions in order to monitor the convergence of computed properties. In this so-called layer-by-layer (LBL) approach, only the new layer of orbitals appearing at each step is optimized, while the previous ones are fixed.

Once the orbitals have been optimized, the quantum electrodynamic (QED) corrections and the Breit interaction are included in the subsequent RCI calculations. In this stage, single and double excitations from the 1$s$ subshell are allowed in order to account for the core-valence (CV) and core-core (CC) electron correlation effects required for achieving accurate results. Meanwhile, larger MR are needed to be considered in order to capture more contributions from triple and quadruple excitations. Here, the additional MR are defined in the following way:

$\rm{MR_1}$$:~1s^2nln'l'~(2\leq n \leq n' \leq 8)$,

$\rm{MR_2}$$:~1s^2nln'l'~(2\leq n \leq n' \leq 9)$,

$\rm{MR_3}$$:~1s^2nln'l'~(2\leq n \leq n' \leq 10)$,

$\rm{MR_4}$$:~1s^2nln'l'~(2\leq n \leq n' \leq 11)$,

$\rm{MR_5}$$:~1s^2nln'l'~(2\leq n \leq n' \leq 12)$,

$\rm{MR_6}$$:~1s^2nln'l'~(2\leq n \leq n' \leq 13)$,

$\rm{MR_7}$$:~1s^2nln'l'~(2\leq n \leq n' \leq 14)$.

\noindent For all MR sets above, the orbital angular momenta follow the standard condition $0 \leq l \leq n-1,~0 \leq l' \leq n'-1$. Additionally, when either principal quantum number $n$ or $n'$ satisfies $n \geq 9$ or $n' \geq 9$, the corresponding orbital angular momenta are restricted to $l,l' \in \{s,p,d,f\}$.

The final MR is chosen as $\rm{MR_7}$. By allowing single and double excitations of all electrons from these configurations to the $\rm{AS}_{15}$ orbital active set, more contributions from triple and quadruple excitations are introduced. The resulting number of CSFs in the final even and odd parity RCI spaces are $129~247~141$ and $118~235~059$, respectively, distributed over the different $J$ symmetries. 
Dealing with such large expansions is challenging in both computer memory and calculation time, and it cannot be handled with the original {\sc Grasp} package. Introducing the configuration state function generators (CSFG)~\citep{Li.2023.p108562} and new condensation methods recently developed as parts of the {\sc Graspg} package~\citep{Si.2025.p109604}, we can effectively reduce the number of CSFs. Briefly, we accumulate the squared weights of the \mbox{CSFGs} from a smaller RCI calculation based on the Dirac-Coulomb Hamiltonian to a specified fraction, as discussed in detail in section 5.3 of~\citet{Li.2023.p108562}. Here we start with $\rm{MR_7}$ and $\rm{AS_9}$ expansions. We perform the condensation by retaining all labelling and generating CSFGs that contribute to 0.9999999 of the accumulated weight to obtain a reduced CSFG list. Then, taking advantage of the features of the CSFGs, we can use a priori condensation of the expansions by redefining the highest orbitals for each symmetry to $\rm{AS_{15}}$. The resulting condensed CSFG list is used for a RCI calculation based on the Dirac-Coulomb-Breit Hamiltonian with limitations on the Breit interaction. The final condensed CSFG lists contain $6~714~901$ and $6~620~286$ CSFs, respectively, for the two parities. Only about 5\% out of the original CSFs are retained. Test calculations show that the significant reduction of the expansion sizes leads to negligible changes of the excitation energies ($\leq 1~\rm{cm^{-1}}$). Given this advantage, we can shorten the entire calculation time, which includes the time for the MCDHF and RCI calculations and the time for evaluating transition rates, lifetimes, hyperfine structures and isotope shifts, to one month using 64 cores of a parallel computer system equipped with two AMD EPYC 7542 central processing units (CPUs).

In the relativistic calculations, $jj$-coupling is always used to label the CSFs, over which the wave functions are expanded. In order to be consistent with the labelling system used by experimentalists or databases such as NIST ASD, $jj$-coupled CSFs are transformed to the $LSJ$-coupled scheme~\citep{Gaigalas.2003.p99}. Meanwhile, the leading term of the $LS$ percentage composition is used to label the quantum states~\citep{Gaigalas.2004.p239,Gaigalas.2017.p6}.

\subsection{Transition rates} \label{subsec:trans}

When a set of atomic state functions has been obtained, transition rates for a multipole transition between two atomic states, $\psi_\alpha(PJM)$ and $\psi_\beta(P'J'M')$, can be expressed in terms of the reduced transition matrix element
\begin{equation}
\langle{{{\psi}_\alpha(PJ)}\Vert\textbf{O}^{\lambda,(k)}\Vert{{\psi}_\beta(P'J')}}\rangle,
\end{equation}
\noindent where $\textbf{O}^{\lambda,(k)}$ is the electromagnetic operator in Coulomb or Babushkin gauge of multipole order ${k}$ (${\lambda}$ = 1 represents electric multipole and ${\lambda}$ = 0 represents magnetic multipole). If needed, transformations of the atomic state functions can be performed to make the orbital sets biorthogonal before Racah algebra is applied \citep{Gaigalas.1997.p3747} to express the transition matrix elements into a sum over one-electron matrix elements~\citep{Olsen.1995.p4499,Joensson.1998.p4967}.

\subsection{Land$\rm{\acute{e}}$ g-factors} \label{subsec:lande}

The Zeeman effect is caused by the interaction between the magnetic moment of the atom and an external magnetic field. Land$\acute{\rm{e}}$ $g$-factors determine the splitting of magnetic sublevels in external magnetic fields. Considering the non-relativistic condition, the Land$\acute{\rm{e}}$ $g$-factor in pure $LS$-coupling can be written as
\begin{equation}
{g_{J}(LS)} = 1+({g_s}-1){\frac{{J(J+1)+S(S+1)-L(L+1)}}{2{J(J+1)}}},
\end{equation}
where $g_s =2.002~319$ is the gyromagnetic ratio for the electron. For low-Z atoms and low-energy states, where relativistic effects can be neglected, or when spin-orbit coupling is weak, the result obtained from the non-relativistic formula is valid.

In fully relativistic theory, the Land$\rm{\acute{e}}$ $g$-factor is given by
\begin{equation}
\begin{aligned}
{g_J} = \frac{2}{\sqrt{{J(J+1)(2J+1)}}}\times\langle & {{{\psi}_\alpha(PJ)}\Vert\sum \limits^{{N}}_{{j=1}} \left[-{\mbox{i}}\frac{\sqrt{2}}{2\alpha}{r_{j}(\bm{\alpha}_{j}\textbf{C}^{(1)}(j))^{(1)}}\right.}  \\ & + {\left.{\frac{g_s-2}{2}\beta_{j}\bm{\Sigma}_{j}}\right]\Vert{{\psi}_\alpha(PJ)}}\rangle,
\end{aligned}
\end{equation}
where $\mbox{i}$ is the imaginary unit, $\alpha$ is the fine-structure constant, $\boldsymbol{\alpha}_j$ is the Dirac matrix, $\textbf{C}^{(k)}$  is a spherical tensor operator and $\boldsymbol{\Sigma}_j$ is the relativistic spin-matrix~\citep{Andersson.2008.p156,Li.2020.p107211}. This formula, including formulas in section~\ref{subsec:hyper}, may appear different in various papers due to different versions of Wigner-Eckart theorem as described in detail in~\citet{Joensson.2022.p7}.

\subsection{Hyperfine structures} \label{subsec:hyper}

The electronic tensor operators for an $N$-electron atom, in atomic units, can be written as~\citep{Andersson.2008.p156,Li.2020.p107211}
\begin{equation}
\textbf{T}^{(1)} = \sum_{j=1}^N - {\mbox{i}} \sqrt{2} \alpha r_j^{-2}(\boldsymbol{\alpha}_j \textbf{C}^{(1)}(j))^{(1)},
\end{equation} 
\begin{equation}
\textbf{T}^{(2)} = \sum_{j=1}^N -r^{-3}_j \textbf{C}^{(2)}(j).
\end{equation}

The hyperfine interaction couples the electronic angular momenta $J$ and the nuclear angular momenta $I$ to a total angular momentum $F = I + J$. The hyperfine interaction constants are defined in atomic units as
\begin{equation}
A_J = \frac{\mu_I}{I} \frac{1}{\sqrt{J(J+1)(2J+1)}} \left \langle {{{\psi}_\alpha(PJ)}} \left \| {\textbf{T}^{(1)}} \right \| {{{\psi}_\alpha(PJ)}} \right \rangle,
\end{equation}
\begin{equation}
B_J = 2Q \sqrt{\frac{J(2J-1)}{(J+1)(2J+1)(2J+3)}} \left \langle {{{\psi}_\alpha(PJ)}} \left \| {\textbf{T}^{(2)}} \right \| {{{\psi}_\alpha(PJ)}} \right \rangle.
\end{equation}
In these expressions, $\mu_I$ is the nuclear magnetic dipole moment and $Q$ is the nuclear electric quadruple moment.

\subsection{Isotope shifts} \label{subsec:iso}

Considering the influence of the finite mass of the nucleus and the spatial distribution of the charge, different isotopes of the same element will have different energy levels. The resulting isotope shift is composed of a mass shift (MS) and a field shift (FS) contributions.

The mass shift is caused by nuclear recoil motion, which can be divided into normal mass shift (NMS) and specific mass shift (SMS). 
The formula for the level isotope shift can be written as follows~\citep{Naze.2013.p2187,Ekman.2019.p433}
\begin{equation}
    \delta \nu_{\rm{MS}} = \left(\frac{M'-M}{MM'}\right)\left(K_{\rm{NMS}}+K_{\rm{SMS}}\right),
\end{equation}
where $M$ and $M'$ are the nuclear masses of different isotopes. $K_{\rm{NMS}}$ and $K_{\rm{SMS}}$ represent mass shift parameters.

The field shift is caused by different isotopes of the same element having different nuclear charge distribution, resulting in small differences in electric potential. This contribution to the level isotope shift can be approximated as
\begin{equation}
    \delta \nu_{\rm{FS}} \approx \left( F^{(0)\rm{ved}} + F^{(1)\rm{ved}} \delta \left \langle r^2 \right \rangle \right) \delta \left \langle r^2 \right \rangle,
\end{equation}
where $F^{(0)\rm{ved}}$ and $F^{(1)\rm{ved}}$ are the relevant level electronic
factors, and the superscript "ved" means "varying electron density" inside the nucleus. $\delta \left \langle r^2 \right \rangle$ is the difference of mean squares of nuclear radii of two isotopes. Details of the level electronic factors can be found in~\citet{Ekman.2019.p433}. Once the level isotope shifts have been computed, the theoretical transition isotope shifts are obtained by subtracting the level isotope shifts for the upper and lower levels in a set of transitions.

\section{Evaluation of Data} \label{sec:floats}

\subsection{Energy levels} \label{subsec:energy}

\begin{figure}
    \centering
    \includegraphics[width=8.5cm]{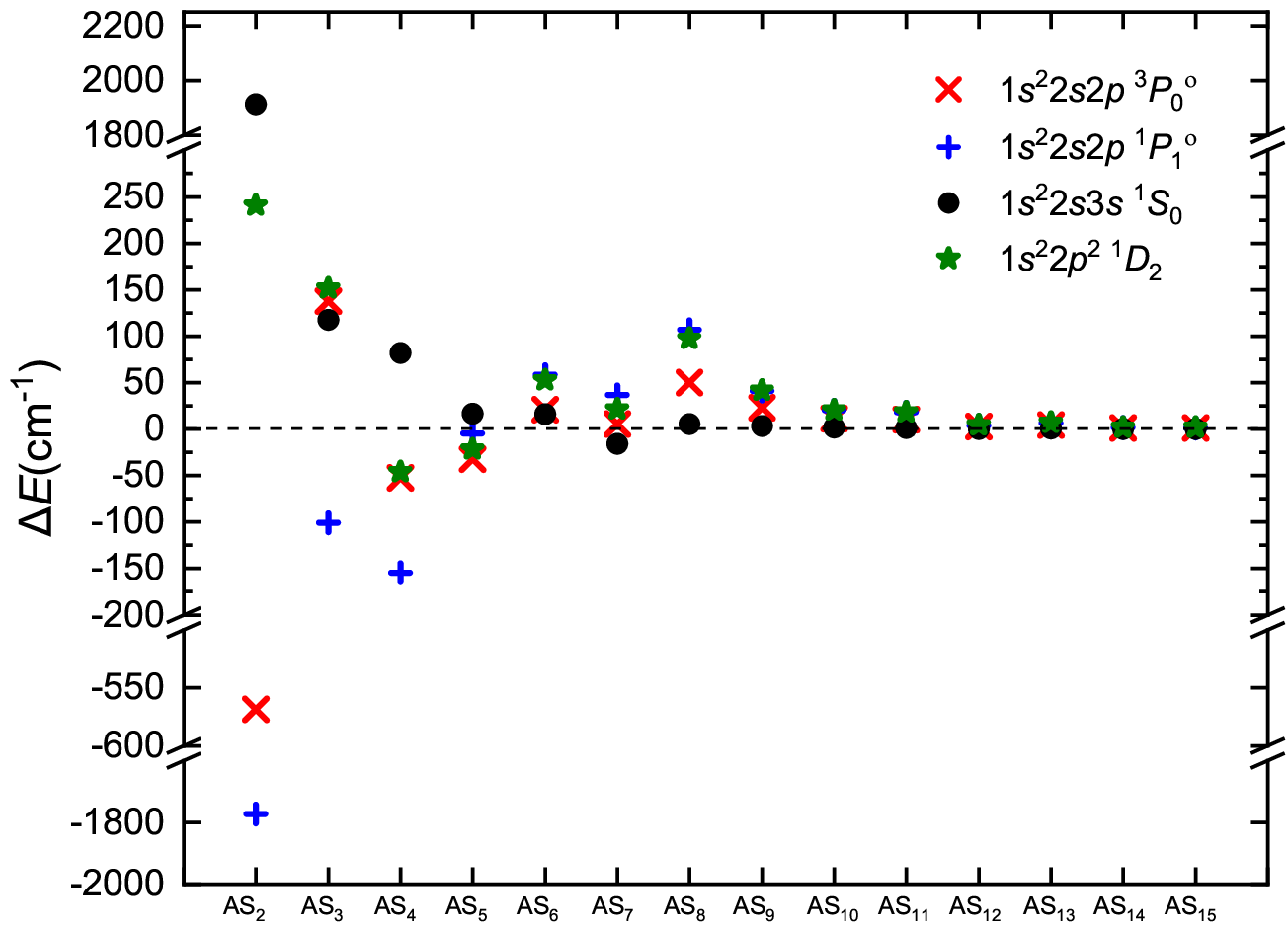}
    \caption{Difference (in $\rm{cm^{-1}}$) of the MCDHF/RCI excitation energies for the $1s^2 2s 2p~^{3}P_{0}^{\circ}$, $1s^2 2s 2p~^{1}P_{1}^{\circ}$, $1s^2 2s 3s~^{1}S_{0}$, $1s^2 2p^2~^{1}D_{2}$ levels as a function of AS. MR is set as $\rm{MR_0}$.}
    \label{fig:fe1}
\end{figure}

In Fig.~\ref{fig:fe1}, four levels ($1s^22s2p~^3P_0^{\circ}$, $1s^22s2p~^1P_1^{\circ}$, $1s^22s3s~^1S_0$, $1s^22p^2~^1D_2$) of Be I are presented as an example to show the effects of the increasing active sets of orbitals on the excitation energies. The MR for the convergence test is $\rm{MR_0}$. $\Delta E$ is the difference of excitation energies obtained with two successive active sets, i.e. $\rm{AS}$$_n$ and $\rm{AS}$$_{n-1}$. It can be clearly seen from Fig.~\ref{fig:fe1} that, with the increasing orbital set, the present calculations are well converged. The differences between adjacent $\rm{AS}$$_n$ are decreased by progressively extending the orbital set. Initially, when the AS is extended from $\rm{AS_{1}}$ to $\rm{AS_{2}}$, the change of the excitation energy is enormous. Especially for $2s2p~^1P_1^{\circ}$ and $2s3s~^1S_0$, the differences are $-$1773 $\rm{cm^{-1}}$ and 1913 $\rm{cm^{-1}}$, respectively. 
Even when the AS is extended to $\rm{AS_{12}}$, the difference of excitation energies between $\rm{AS_{12}}$ and $\rm{AS_{13}}$ is 4.3 $\rm{cm^{-1}}$ for $1s^22s2p~^3P_0^{\circ}$, 6.4 $\rm{cm^{-1}}$ for $1s^22s2p~^1P_1^{\circ}$ and 7.0 $\rm{cm^{-1}}$ for $1s^22p^2~^1D_2$. Not until the expansion of AS reaches $\rm{AS_{15}}$, the difference of excitation energies falls below 1 $\rm{cm^{-1}}$. Therefore, it is necessary to expand the AS to such a large extent in order to reach spectroscopic accuracy.

In the present RCI calculations, we have constructed  sufficiently large CSF expansions and fully considered various correlation effects. In view of the fact that Be I only have four electrons, VV, CV and CC electron correlation effects are all included. The corresponding average contributions to the excitation energies are 169 $\rm{cm^{-1}}$, 94 $\rm{cm^{-1}}$ and 159 $\rm{cm^{-1}}$ respectively. Additionally, the Breit contributions between CSFs built by orbitals with $n \leq 11,~l \leq 5$ and the leading QED corrections are also considered, the average contributions of which to excitation energies are only 2.4 $\rm{cm^{-1}}$ and 1.8 $\rm{cm^{-1}}$, respectively.

\begin{figure}
    \centering
    \includegraphics[width=8.5cm]{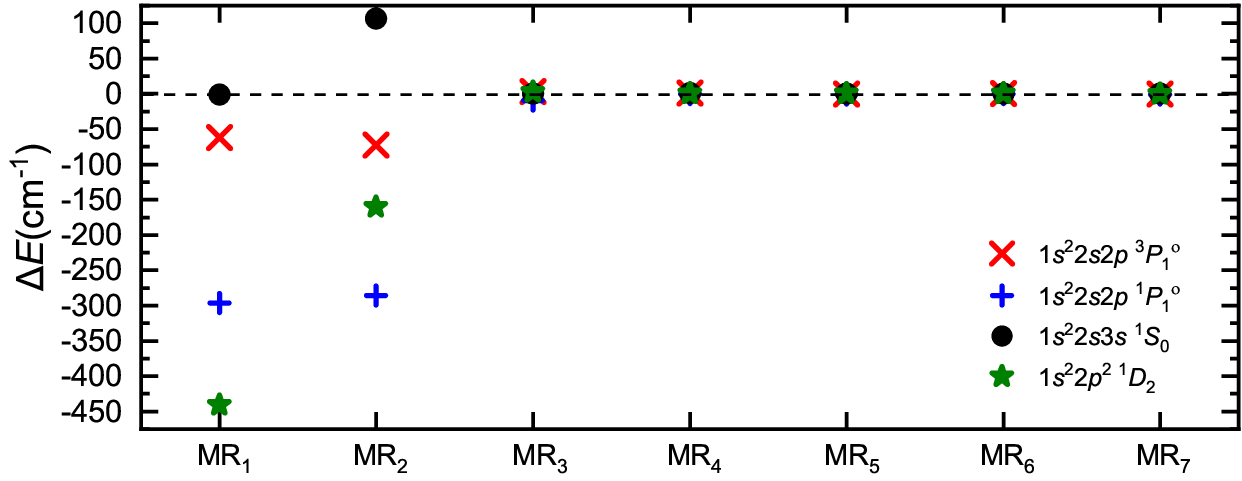}
    \caption{Difference (in $\rm{cm^{-1}}$) of the MCDHF/RCI excitation energies for the $1s^2 2s 2p~^{3}P_{0}^{\circ}$, $1s^2 2s 2p~^{1}P_{1}^{\circ}$, $1s^2 2s 3s~^{1}S_{0}$, $1s^2 2p^2~^{1}D_{2}$ levels as a function of MR. AS is set as $\rm{AS_{15}}$.}
    \label{fig:fe2}
\end{figure}

Furthermore, it is worth noting that, when extending the MR spaces for the final RCI calculations, the energies can also exhibit considerable changes. Effects of enlarging the MR on excitation energies for $1s^22s2p~^3P_0^{\circ}$, $1s^22s2p~^1P_1^{\circ}$, $1s^22s3s~^1S_0$, $1s^22p^2~^1D_2$ in $\rm{cm^{-1}}$ are shown in Fig.~\ref{fig:fe2}. The AS of orbitals is set as $\rm{AS_{15}}$ and here $\Delta E$ is the difference of excitation energies obtained with $\rm{MR}$$_n$ and $\rm{MR}$$_{n-1}$. Inspection of Fig.~\ref{fig:fe2} clearly illustrates that the triple and quadruple excitations introduced through additional configurations in the MR can greatly influence the excitation energies. The excitation energies are generally lowered by extending the MR from $\rm{MR_0}$ to $\rm{MR_1}$. For the levels $1s^22p^2~^1D_2$, $1s^22s2p~^1P_1^{\circ}$ and $1s^22s2p~^3P_0^{\circ}$, $\rm{MR_1}$ decreases the energies by about 441 $\rm{cm^{-1}}$, 296 $\rm{cm^{-1}}$ and 62 $\rm{cm^{-1}}$. When it comes to $\rm{MR_2}$, the newly introduced MR can raise or lower the excitation energies. For the levels $1s^22p^2~^1D_2$, $1s^22s2p~^1P_1^{\circ}$ and $1s^22s2p~^3P_0^{\circ}$, $\rm{MR_2}$ produces similar effect as $\rm{MR_1}$, while for the level $1s^22s3s~^1S_0$, the excitation energy is increased by about 107 $\rm{cm^{-1}}$. Then, keeping enlarging MR has no further sizable effect on the excitation energies given that these effects are almost below 1 $\rm{cm^{-1}}$. However, as will be shown in section~\ref{subsec:tran}, additional triple and quadruple excitations are very important for the transition rates, especially for improving the agreement between different gauges.

In Table~\ref{tab:Level}, the excitation energies ($E_{\rm{MCDHF}}$) of the 99 lowest levels of configurations $1s^2 2s nl$ ($n \leq 7$) + $1s^2 2p^2$ in Be I from the present RCI calculations are provided. The experimental values from the NIST ASD~\citep{Kramida.2023.p} are also listed in the table. The configuration, total angular number $J$ and energy ordering are adopted as the good quantum numbers to match the levels. It can be seen from Table~\ref{tab:Level}, NIST ASD provides 75 $n \leq 7$ levels for Be~I. The present MCDHF values agree very well with these observations, and the differences are all within 10~$\rm{cm^{-1}}$, which is highly satisfactory. The total average difference with the standard deviation from the NIST values is 7.08 $\pm$ 1.14 $\rm{cm^{-1}}$ (0.011$\%$ $\pm$ 0.003$\%$). It is worth noting that the standard deviation is quite small, which means that if we correct the ground-state energy of Be~I by shifting it down by 7.08 $\rm{cm^{-1}}$, the present excitation energies agree with the experimental values with a root-mean-square difference of 1.2~$\rm{cm^{-1}}$. To quantify shifted absolute energy differences, the mean level deviation (MLD) which has been described in~\citet{Ekman.2014.p24} can also be used.

\begin{equation}
MLD = \frac{1}{N}\sum_{i=1}^{N}|E_{obs}(i)-E_{calc}(i)+ES|,
\end{equation}

\noindent where NIST values are used as $E_{obs}$. The energy shift (ES) is chosen as to minimize the sum and reflects the degree to which the ground state level is either favored or disfavored in the theoretical balance of binding energy. The obtained MLD (ES) value is 0.75 (6.91) $\rm{cm^{-1}}$ for this work.

As we have discussed in section~\ref{subsec:mcdhf}, the condensation method may change the computed excitation energies by up to 1~$\rm{cm^{-1}}$. For this reason, $ab~initio$ calculated excitation energies ($E_{\rm{MCDHF}}$) are rounded to integer numbers in units of $\rm{cm^{-1}}$. Additionally, the excitation energies which are used for subsequent calculations are corrected with the help of experimental results. For energy levels which have no reference to experimental values (involving terms$~^1G$,$~^3G$,$~^1H^{\circ}$,$~^3H^{\circ}$,$~^1I$,$~^3I$), we use the differences between the present MCDHF results of terms$~^1P^{\circ}$,$~^3P^{\circ}$,$~^1D$,$~^3D$,$~^1F^{\circ}$,$~^3F^{\circ}$ and the NIST values to extrapolate the approximate values, i.e., increasing the original MCDHF values by 7.08 $\rm{cm^{-1}}$. Retaining two digits after the decimal point, these adjusted values, which are labeled as $E_{\rm{adjusted}}$, are also listed in Table~\ref{tab:Level}. 

For the beryllium atom, excitation energies of $n \leq 3$ levels were calculated by ~\citet{FroeseFischer.2004.p1} using the MCHF+BP method. The average deviation between their results and experimental values is around 100 $\rm{cm^{-1}}$. The differences between the results of~\citet{Dong.2021.p43103}, who also used the MCDHF and RCI method, and the NIST values can reach 4\% while our results are two orders of magnitude better. In recent years, the most accurate results for Be I have been obtained with the all-electron ECG functions by Stanke $et~
al.$~\citep{Stanke.2007.p,Stanke.2007.pa,Stanke.2009.p,Stanke.2018.p,Stanke.2019.p,Stanke.2019.p155002,Stanke.2021.p138823,Stanke.2022.p,Kedziorski.2020.p137476} and Nasiri $et~al.$~\cite{Nasiri.2021.p}. Their calculations involved $^1S$, $^1P^{\circ}$, $^1D$ and $^3P^{\circ}$ terms of $n \leq 7$ levels. All-electron explicitly correlated Gaussian functions and a finite-nuclear-mass variational method were used in their calculations and the energies of the states are augmented with the leading relativistic and QED corrections. As a result, the deviation between their results and experimental values is commonly less than 0.1 $\rm{cm^{-1}}$, which is highly satisfactory. To reach such a benchmark quality, this approach requires about one year of continuous computing with hundreds of cores on a parallel computer system.
For the present MCDHF/RCI calculations, calculation time and computational load can be greatly reduced. The present energy data are generally accurate enough and more complete than existing theoretical data, and can be useful for astrophysical research.

\subsection{Transition Rates} \label{subsec:tran}

In Table~\ref{tab:Tr}, the calculated wavelengths, transition rates, weighted oscillator strengths, line strengths and branching fractions (BF) for the E1, M1, E2 and M2 transitions with BF $\geq~10^{-5}$ among the 99 lowest levels of configurations $1s^2 2s nl$ ($n \leq 7$) + $1s^2 2p^2$ in Be~I are presented. Here, for the E1 and E2 transitions, both the results in the Babushkin (length) form and Coulomb (velocity) form are provided. Since there are only fifteen E2 transitions with BF $\geq~10^{-5}$ in this work, the subsequent uncertainty estimations primarily focus on E1 transitions.

\begin{figure}
    \centering
    \includegraphics[width=8.5cm]{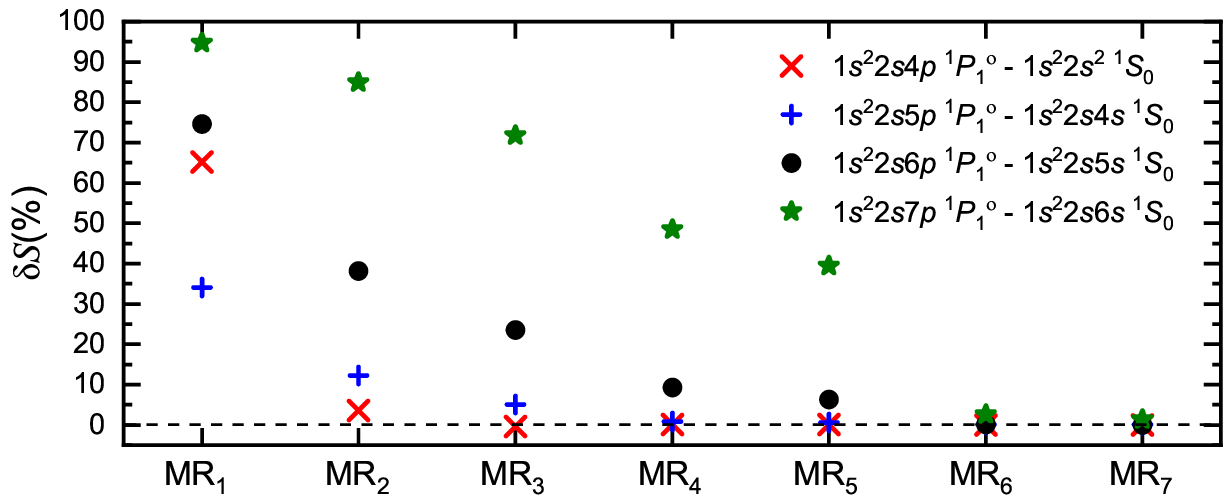}
    \caption{Effects (in $\rm{cm^{-1}}$) of extending the MR on $\delta S$ for the transitions of $1s^2 2s 4p~^{1}P_{1}^{\circ} - 1s^2 2s^2~^{1}S_{0}$, $1s^2 2s 5p~^{1}P_{1}^{\circ} - 1s^2 2s 4s~^{1}S_{0}$, $1s^2 2s 6p~^{1}P_{1}^{\circ} - 1s^2 2s 5s~^{1}S_{0}$, $1s^2 2s 7p~^{1}P_{1}^{\circ} - 1s^2 2s 6s~^{1}S_{0}$. AS is set as $\rm{AS_{15}}$.}
    \label{fig:ft1}
\end{figure}

The agreement between the two gauges is indeed often used as a good accuracy indicator~\citep{FroeseFischer.2009.p14019,Ekman.2014.p215}. The difference $\delta S$ between line strength $S_l$ and $S_v$ (in length and velocity forms respectively) is defined as $\delta S$ = $\left|S_{v} - S_{l} \right|$/$\min (S_{v}$,~$S_{l})$. It needs to be mentioned that, as observed in section~\ref{subsec:energy}, the difference $\delta S$ is also affected when extending the MR spaces for the final RCI calculations. Fig.~\ref{fig:ft1} shows the effects of MR on $\delta S$ for four strong transitions, i.e. $1s^2 2s 4p~^{1}P_{1}^{\circ} \rightarrow 1s^2 2s^2~^{1}S_{0}$, $1s^2 2s 5p~^{1}P_{1}^{\circ} \rightarrow 1s^2 2s 4s~^{1}S_{0}$, $1s^2 2s 6p~^{1}P_{1}^{\circ} \rightarrow 1s^2 2s 5s~^{1}S_{0}$, $1s^2 2s 7p~^{1}P_{1}^{\circ} \rightarrow 1s^2 2s 6s~^{1}S_{0}$. It is obvious that, with the expansion of MR, the $\delta S$ values have all converged for these transitions. For the transition  $1s^2 2s 7p~^{1}P_{1}^{\circ} \rightarrow 1s^2 2s 6s~^{1}S_{0}$, note that $\delta S$ drops from 39\% to 2.6\% when MR is extended from $\rm{MR_5}$ to $\rm{MR_6}$. This change is due to the irregularity of the mean radii of the correlation orbitals which are usually relatively small, and strongly dependent on the optimization strategy used for obtaining the wave function~\citep{Papoulia.2019.p106}. The correlation orbitals are usually rather contracted and the energetically dominant effects must be first saturated to obtain orbitals localized in other regions of space. However, for the wave function used in present work, the mean radius of $13s$ orbital is much larger than that of $12s$ or $11s$. Therefore, when $13s$ is included in the MR, $\delta S$ exhibits a considerable change. The line strengths in both gauges change, but consistent with the analysis in \citet{Papoulia.2019.p106}, where the line strength in the length gauge, $S_l$, changes the most.

Using the uncertainty estimation method suggested by~\citet{Kramida.2013.p313,Kramida.2014.p86,Kramida.2024.p}, for E1 and E2 transitions we provide the estimated uncertainties of line strengths $S$ adopting the NIST ASD~\citep{Kramida.2023.p} terminology (AA $\leq$ 1~\%, A$^{+}$ $\leq$ 2~\%, A $\leq$ 3~\%, B$^{+}$ $\leq$ 7~\%, B $\leq$ 10~\%, C$^{+}$  $\leq$ 18~\%,  C $\leq$ 25~\%,  D$^{+}$ $\leq$ 40~\%, D $\leq$ 50~\%, and E $>$ 50~\% ) in the penultimate column of Table~\ref{tab:Tr} (labelled as $\rm{Acc_1}$). Following the approach outlined in~\citet{Kramida.2024.p}, line strength serves as the discriminating quantity for transitions from $n \leq 4$, while branching fraction is used for those from $n = 5,~6,~7$. The final estimate of the fractional uncertainty of $S$ is taken as the maximum of the following three values:

1. The value of exp(exp|$u(Q)$|)-1, where $u(Q)$ is the fitting function based on the discriminating quantity $Q$. The function $u(Q)$ is obtained by first partitioning the data based on $Q$. For each partition, the average $Q$ is calculated and paired with the root-mean-square value of $\text{ln}(S_L/S_V)$ computed over that specific range of $Q$. A curve is then fitted to these resulting data pairs. For $n \leq 4$ transitions, the fitting function $u(Q)$ can be written as: $u(Q)=0.00564Q^2-0.24023Q-5.94218$, where $Q \equiv \text{ln}(<S_L>)$;
For $n = 5$ transitions: $u(Q) = 0.00695Q^2 - 0.27196Q - 7.48159$;
For $n = 6$ transitions: $u(Q) = 0.00801Q^2 - 0.11431Q - 5.10159$;
For $n = 7$ transitions: $u(Q) = 0.0096Q^2 - 0.00932Q - 3.41755$.
In the last three cases ($n \geq 5$), $Q$ corresponds to $\text{ln}(<BF>)$. Across all cases, $u(Q)$ corresponds to $\text{ln}(\sqrt{<(\text{ln}(S_\text{L}/S_\text{V}))^2>})$.

2. The actual value of exp(|ln($S_\text{L}/S_\text{V}$)|)-1.

3. The minimum error intrinsic to the MCDHF method. Several sources of uncertainty are identified. First, the average numerical convergence error, obtained from calculations using $\text{MR}_7$ with $\text{AS}_{14}$ and $\text{AS}_{15}$, is 0.48$\%$. Second, higher-order relativistic and QED contributions are expected to be well below 0.1$\%$ for this atomic system. Third, the computational approximation due to the reduction of the CSF expansion is taken into account. Its impact on energy levels (within 1 $\text{cm}^{-1}$) corresponds to an uncertainty of 0.01$\%$ in the transition probabilities. After comprehensive consideration, we conclude that the minimum intrinsic error is estimated to be 0.5$\%$.

\begin{figure}
    \centering
    \includegraphics[width=8.5cm]{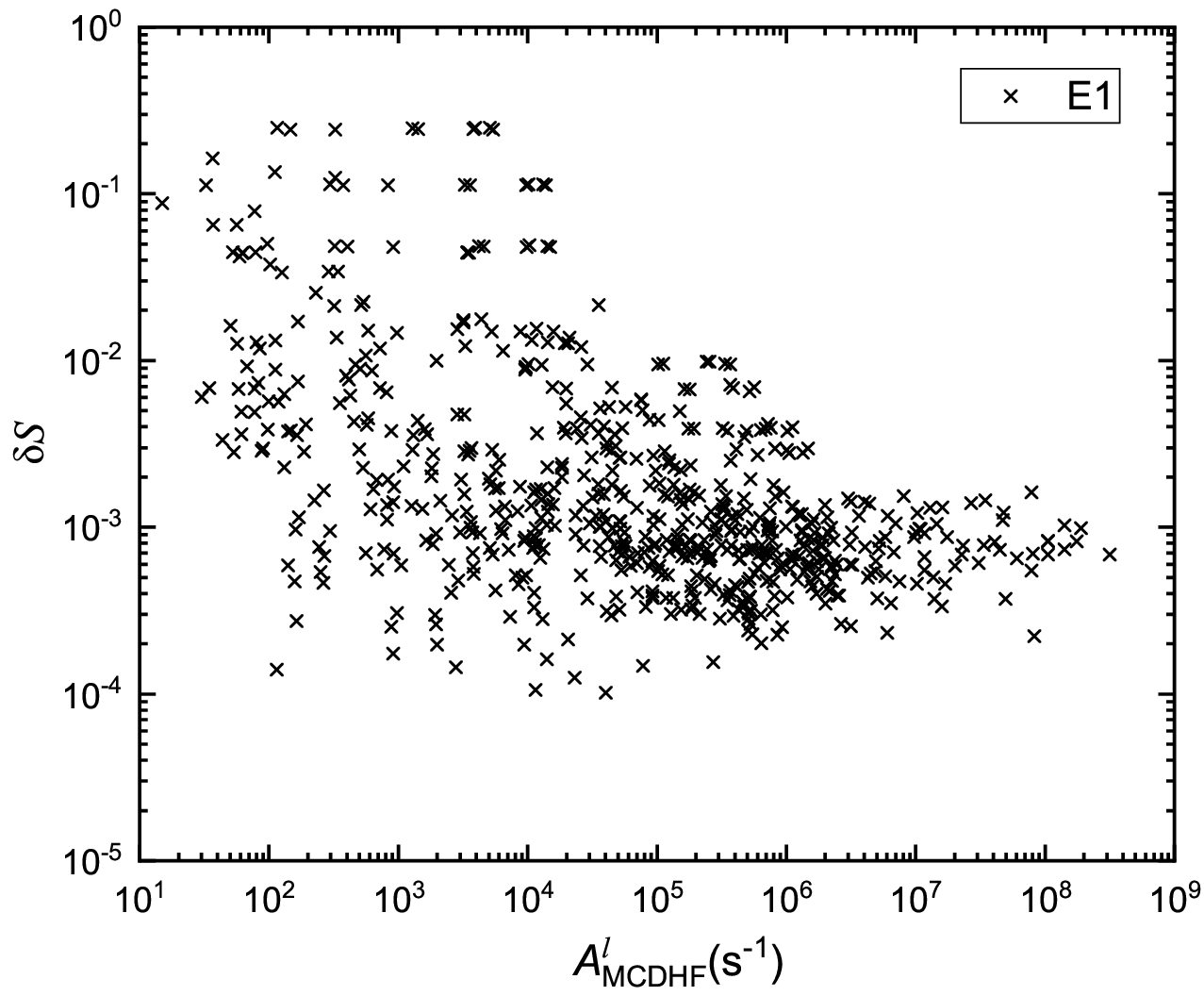}
    \caption{The difference $\delta S$ for E1 transitions with BF $\geq 10^{-5}$ among the lowest 99 $n \leq 7$ levels.}  
    \label{fig:Sal}
\end{figure}

This process is applied to the E1 and E2 transitions of Be I. In Table~\ref{tab:Tr}, about 36\% of E1 $S$ values in Be I are evaluated to have uncertainties of $\leq$ 1~\% (AA), 12~\% have uncertainties of  $\leq$ 2~\% (A+), 4~\% have uncertainties of  $\leq$ 3~\% (A), 37~\% have uncertainties of  $\leq$ 7~\% (B+), 6~\% have uncertainties of  $\leq$ 10~\% (B) while only 5~\% have uncertainties of  $>$ 10~\% (C+, C, D+, D and E). For the only M2 transition $2s\,2p~^{3}P_{2}^{\circ} \rightarrow 2s^{2}~^{1}S_{0}$ in Table~\ref{tab:Tr}, owing to the absence of a comparison between the length and velocity gauges, its accuracy is estimated by comparing the results in two consecutive layers of the calculation. Based on $\text{MR}_7$, we employ the results from $\text{AS}_{14}$ and $\text{AS}_{15}$ for comparison. The resulting line strength difference is less than $0.01~\%$, leading to an AA rating for this transition. Fig.~\ref{fig:Sal} presents the distribution of difference $\delta S$ as a function of the transition rates for E1 transitions with BF $\geq~10^{-5}$ and it's evident that strong transitions have smaller $\delta S$.

\begin{figure}
    \centering
    \includegraphics[width=10cm]{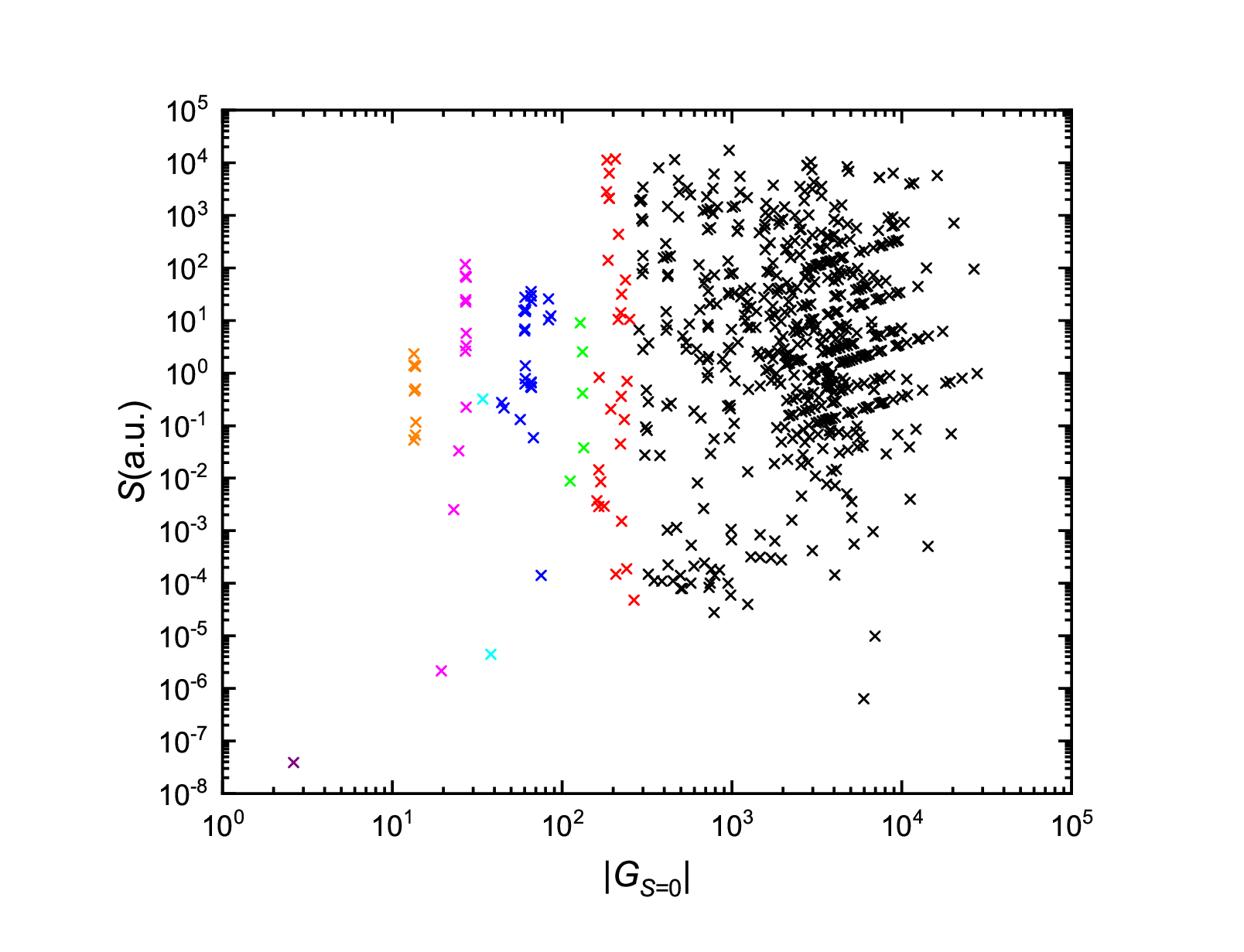}
    \caption{Distribution of line strength $S$ as a function of the $G_{S=0}$ parameter for E1 transitions with BF $\geq 10^{-5}$ among the 99 levels. Different colours mark the accuracy class of estimated transitions based on the methodology described in~\citet{Rynkun.2022.p82}. Black colour means AA accuracy class, red - A+, green - A, blue - B+, cyan - B, magenta - C+, orange - C, purple - E.} 
    \label{fig:Ges}
\end{figure}

The accuracy of the wave functions and the transition parameters is also evaluated by the quantitative and qualitative evaluation approach, which studies the relation between the line strength $S$ and the gauge parameter $G$~\citep{Grant.1974.p1458,Kaniauskas.1974.p463}. The uncertainty estimation methodology is described in detail by~\citet{Rynkun.2022.p82} and~\citet{Kitoviene.2024.p}. The dependence of the line strength $S$ on $G$ is parabolic, and the minimum of the parabola for E1 transitions can be expressed as follows

\begin{equation}
{G_{S=0}} = {\frac{\sqrt{2}}{1-\text{sgn}(\frac{M_l}{M_v})\sqrt{\frac{S_l}{S_v}}}},
\end{equation}

\noindent where $\text{sgn}\frac{M_l}{M_v}$ shows the sign of reduced transition matrix elements. In this approach, we need to focus on the signs of the reduced matrix element $M_l$ (in the Babushkin form) and the reduced matrix element $M_v$ (in the Coulomb form). If the signs of the reduced matrix elements in the two forms are different, the range of the $G_{S=0}$ parameter is $0<G_{S=0}<\sqrt{2}$. As it was mentioned in~\citet{Rynkun.2022.p82}, transitions belonging to this range are not evaluated and are assigned to the accuracy class E directly. If the reduced matrix elements have the same sign in the two forms, then we can use $G_{S=0}$ to estimate the accuracy class. The connection of the limits of $G_{S=0}$ for the accuracy classes in the case of the reduced matrix elements with the same sign are presented in~\citet{Gaigalas.2022.p281}. The distribution of line strength $S$ as a function of $G_{S=0}$ parameter for E1 transitions with BF $\geq 10^{-5}$ among 99 levels is shown in Fig.~\ref{fig:Ges}. Most of the transitions belong to AA accuracy class. These estimated uncertainties of line strengths $S$ are listed in the last column of Table~\ref{tab:Tr} (labelled as $\rm{Acc_2}$). About 89\% have uncertainties of $\leq$ 1\% (AA), 4\% have uncertainties of $\leq$ 3\% (A), and only 7\% have uncertainties of $\geq$~3\%. It is obvious that uncertainties given by the quantitative and qualitative evaluation approach are smaller and more optimistic while those given by the method suggested by Kramida tend to be larger and less optimistic due to introduction of the averaged uncertainties for the discriminating quantity $Q$ in various ranges of $Q$. To be safe, we recommend the readers to use the more conservative estimates of the $\text{Acc}_1$ column of Table~\ref{tab:Tr}, although we think that $\text{Acc}_2$ may be adequate for most transitions. This topic still needs further research.

To illustrate the accuracy of the calculated transition data of this work, the present results from bound excited singlet $S$ and $P^{\circ}$ states are compared with available theoretical oscillator strengths in Table~\ref{tab:TrC}. It can be seen that, for all these strong transitions, the deviation between oscillator strength in the Babushkin form and the Coulomb form are less than 1\%. Also comforting is the excellent agreement between our MCDHF/RCI oscillator strengths and the ECG results~\citep{Nasiri.2021.p}. The relative differences are indeed within 2\%, which is highly satisfactory, except for 4 transitions (4\% for $2s\,4p~^{1}P_{1}^{\circ} \rightarrow 2s^{2}~^{1}S_{0}$, 2\% for $2s\,5p~^{1}P_{1}^{\circ} \rightarrow 2s\,4s~^{1}S_{0}$, 4\% for $2s\,6p~^{1}P_{1}^{\circ} \rightarrow 2s\,5s~^{1}S_{0}$, 9\% for $2s\,7p~^{1}P_{1}^{\circ} \rightarrow 2s\,6s~^{1}S_{0}$). For all transitions, we investigate their cancellation factors (CF), which reveal large cancellation in the computation of transition parameters~\citep{Gaigalas.2020.p13}, and they are also listed in Table~\ref{tab:Tr}. For these four transitions, their CFs are below 0.01, indicating that the calculated transition parameters are affected by a strong cancellation effect, which is often associated with large uncertainties. It should be noted that for these four transitions with CF < 0.01, the relative discrepancies between the present results and the ECG results exceed the accuracy provided by $\text{Acc}_1$ in three cases, and surpass the accuracy provided by $\text{Acc}_2$ in all four cases. This indicates that the uncertainties given in this work for transitions with CF < 0.01 are likely underestimated and should be used with caution. Apart from the ECG results,~\citet{Chen.1998.p4523} calculated the oscillator strengths of bound states of Be using the B-spline CI with semi-empirical core potential (BCICP) method. Compared with the present MCDHF/RCI values, their results have a larger deviation from those of ECG and some differences exceed 10\%, i.e. $2s\,4p~^{1}P_{1}^{\circ} \rightarrow 2s^{2}~^{1}S_{0}$, $2s\,5p~^{1}P_{1}^{\circ} \rightarrow 2s\,4s~^{1}S_{0}$, $2s\,6p~^{1}P_{1}^{\circ} \rightarrow 2s\,5s~^{1}S_{0}$, $2s\,7p~^{1}P_{1}^{\circ} \rightarrow 2s\,6s~^{1}S_{0}$, the largest difference reaching 46\% for the last transition. 

The experimental and theoretical transition data for Be I before 2010 were evaluated and compiled by NIST~\citep{Fuhr.2010.p}. The main transition results come from two sophisticated MCHF calculations by~\citet{Tachiev.1999.p5805}, and by~\citet{Nam.2002.p}. Approximately 70\% of the transition rates belong to the accuracy class B, with the rest distributed among A, C and D. By comparing 264 transition rates from NIST with our calculated results, we find that many transitions exhibit discrepancies exceeding 5\%. Notably, for transitions originally classified as grade D in the NIST compilation, deviations typically exceed 10\%, with some differing by orders of magnitude. The two earlier MCHF calculations were performed with relatively limited expansion of active space and multireference. In contrast, the computational procedure adopted in the present work is more complete. As discussed in the beginning of section~\ref{subsec:tran}, triple and quadruple excitations introduced through the multireference expansion have a significant impact on the transition rates. The only M2 transition with a notable discrepancy also demonstrates an excellent convergence behavior in our calculation. Therefore, we have reasonable confidence in the improved reliability and accuracy of the transition results obtained in this work. In addition to the 264 transitions rated by NIST, we provide reliability assessments for nearly 500 additional transitions. Given the substantial deviations observed in certain cases and the newly evaluated transitions, we believe that our results have largely enriched the reference data for Be I and could serve as a useful reference for relevant studies.

\subsection{Lifetimes}\label{subsec:life}

Radiative lifetimes $\tau_{\rm{MCDHF}}^l$ in the Babushkin (length) gauge from the present MCDHF/RCI calculations are provided in Table~\ref{tab:Level} for the 99 lowest levels of configurations $1s^2 2s nl$ ($n \leq 7$) + $1s^2 2p^2$ in Be I, which are calculated by including all possible E1, E2, M1 and M2 radiative transition rates for the decay channels. Radiative lifetimes in the Coulomb (velocity) gauge are not provided here because the lifetimes obtained in the length and velocity gauges agree very well. Except for the $2s\,2p~^{3}P_{1}^{\circ}$ level, all other lifetimes agree within 1\%. Instead, we provide the relative uncertainty of the lifetimes calculated in the length gauge. It can be easily determined using the standard formula for error propagation~\citep{Kramida.2024.p}:

\begin{equation}
{\frac{u(\tau_{k})}{\tau_{k}} = \tau_{k}\sqrt{\sum_{i}u(A_{ki})^2}},
\end{equation}

\noindent where $A_{ki}$ means the transition from the upper level $k$ to the lower level $i$ and the summation goes over all transitions to lower levels. The values of $u(A_{ki})$ used here are taken from $\text{Acc}_1$. It can be seen that for levels with $n \leq 6$, the relative uncertainties of the lifetimes are all within 1\%, whereas for the $n = 7$ levels the uncertainties are somewhat larger.

\begin{figure}
    \centering
    \includegraphics[width=8cm]{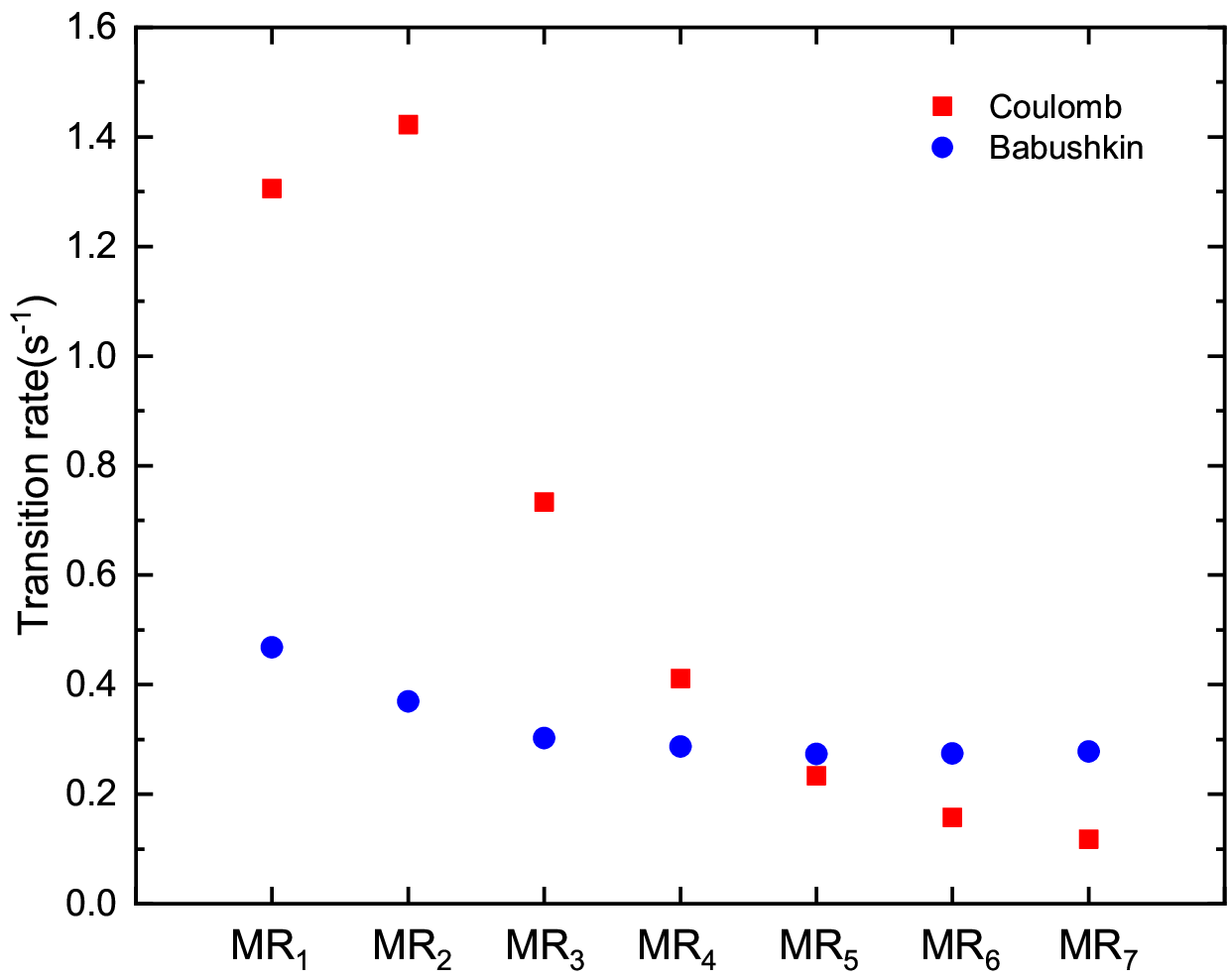}
    \caption{Comparison of the MCDHF/RCI transition rate for the $2s\,2p~^{3}P_{1}^{\circ}~\rightarrow~2s^{2}~^{1}S_{0}$ level as a function of MR for two gauges. AS is set as $\rm{AS_{15}}$.}
    \label{fig:icmr}
\end{figure}

For all levels considered here, the contributions from forbidden transitions are negligibly small, except for two levels ($2s\,2p~^{3}P_{1}^{\circ}$ and $2p^{2}~^{1}D_{2}$). They are also the only two levels with $n \leq 6$ whose lifetime uncertainties exceed 1\%. For $2p^{2}~^{1}D_{2}$, the dominant contribution to its lifetime comes from the $2p^{2}~^{1}D_{2}~\rightarrow~2s\,2p~^{1}P_{1}^{\circ}$ E1 transition, but the 
$2p^{2}~^{1}D_{2}~\rightarrow~2s^{2}~^{1}S_{0}$ E2 transition also play a non-negligible role.
The lifetime for $2s\,2p~^{3}P_{1}^{\circ}$
is mainly determined by the $2s\,2p~^{3}P_{1}^{\circ}~\rightarrow~2s^{2}~^{1}S_{0}$ transition, which is an intercombination transition. In Fig.~\ref{fig:icmr}, we investigate the relationship between the MR and the agreement of transition rates 
evaluated in the two gauges for this transition. This figure clearly shows that, with expansion of the MR, the transition rate in the Babushkin gauge is well converged, while the transition rate in the Coulomb gauge changes dramatically. The $\rm{MR_2}$ and $\rm{MR_7}$ transition rates calculated in the Coulomb gauge differ by one order of magnitude.
This transition is quite sensitive to electron correlation and agreement in the gauges could be accidental. 
Therefore, length-gauge result, which is well converged, appears to be more reliable and should be used. Furthermore, lifetime for $2s\,2p~^{3}P_{1}^{\circ}$ presented in this work still needs further research in view of the discussion about contributions from negative continuum states to the lifetime in the length and velocity gauges~\citep{Chen_2001}.

Comparisons with previous experimental and theoretical lifetimes are provided in Table~\ref{tab:Lif}, where 13 levels for which measured lifetimes are known in the literature are presented~\citep{Bergstroem.1969.p721,Bromander.1969.p523,Andersen.1969.p76,Bromander.1971.p61,Andersen.1971.p52,Hontzeas.1972.p55,Poulsen.1975.p1393,Kerkhoff.1980.p11,Irving.1999.p137,Schnabel.2000.p,Moccia.1985.p3537,Chang.1990.p4922,Fischer.1997.p3333,Wang.2018.p112}. The measurement uncertainties consist of possible remaining systematic errors and statistical errors from different recordings. The experimental lifetimes are highlighted in boldface when the present MCDHF/RCI values fall within the experimental error. For $2s\,2p~^{1}P^{\circ}$, $2s\,3s~^{3}S$, $2p^2~^{3}P$, $2s\,3d~^{3}D$, $2s\,3d~^{1}D$, $2s\,4s~^{3}S$, $2s\,4d~^{3}D$, $2s\,5s~^{3}S$, $2s\,5d~^{3}D$ and $2s\,6s~^{3}S$, the results of this work are in good agreement with many measured and calculated values. Especially for the five new values measured recently by~\citet{Wang.2018.p112} through the time-resolved laser-induced fluorescence technique, all our values fall within the experimental error bars, except for $2s\,6s~^{3}S$, which shows a slight deviation. For $2s\,5d~^{1}D$ and $2s\,6d~^{1}D$, 
the theory-observation differences are large. However, the relevant experimental measurements are relatively old and scarce for these two levels. Despite deviation from experimental measurements, the present MCDHF/RCI values for these two levels are consistent with the theoretical values of~\citet{Moccia.1985.p3537} and~\citet{Chang.1990.p4922}. 

\subsection{Land${\acute{\text{e}}}$ $g$-factors}\label{subsec:lande2}

Table~\ref{tab:HpI} display the Land${\acute{\rm{e}}}$ $g$-factors and their $LS$-coupling values for the lowest 99 levels in Be I. Since the MCDHF method delivers Land${\acute{\rm{e}}}$ $g$-factors as single values without inherent uncertainty estimates, we evaluated their numerical convergence to assess the associated uncertainty. Specifically, we compared results obtained from $\text{AS}_{14}$ and $\text{AS}_{15}$ based on $\text{MR}_7$. The observed variations provide an estimate of the numerical uncertainty inherent in our computational setup and the average uncertainty is 0.01$\%$. It is crucial to distinguish this from the overall accuracy of the method. The total uncertainty is dominated by the systematic bias intrinsic to the MCDHF approach. Therefore, the convergence test primarily validates the stability of our specific calculation rather than its absolute accuracy against a benchmark, and it applies to the subsequent evaluation of uncertainties in the hyperfine structure and isotope shifts.

It is clear that most of the levels have results that are consistent with the pure $LS$-coupling value, meaning that the mixing between terms is either small or occurs between terms with nearly the same $g_J$ values. Taking level~8 as an example, this level is labelled as $^1D_2$, which has a composition of 52\% $2p^{2}~^{1}D_{2}$, 27\% $2s\,3d~^{1}D_{2}$ and 6\% $2s\,8d~^{1}D_{2}$. The pure $LS$-coupling values are all 1.0000 for these three states. The resulting value of the mixed state is therefore identical to the one of the pure $LS$-coupling value. Strong mixing can significantly affect Land${\acute{\rm{e}}}$ $g$-factors if and only if the mixed states show large differences in their $g_J$ value. As shown by~\citet{Schiffmann.2021.p186},~\citet{Li.2020.p25} and~\citet{Wu.2024.p108907}, large deviations from $LS$-coupling are more prominent for heavier ions, where relativistic effects are stronger.

\subsection{Hyperfine structure}\label{subsec:hyperc}

The hyperfine interaction constants among the lowest 99 levels in $^9$Be are presented in Table~\ref{tab:HpI}. For the present calculations, we adopted the following nuclear values: nuclear spin $I$ = 1.5, nuclear magnetic dipole moment $\mu_I$ = -1.177432(3)~\citep{Stone.2015.p} and nuclear electric quadrupole moment $Q$ = 0.05350(14) barns~\citep{Puchalski.2021.p}. The average numerical uncertainties of hyperfine interaction constants are also 
estimated by comparing results from $\text{AS}_{14}$ and $\text{AS}_{15}$ based on $\text{MR}_7$. For $\text{A}_{J}$, the average uncertainty is 0.23$\%$. Results with absolute values below 5 MHz are subject to greater uncertainties. For $\text{B}_{J}$, the average uncertainty is 1.5$\%$ and the uncertainty increases for results with absolute values below 0.002 MHz.

Considering that the $1s^2\,2s\,2p~^{3}P_{1,2}^{\circ}$ states are the lowest excited states in which hyperfine effects can occur, they have been often investigated through both experimental and theoretical approaches over the past few decades~\citep{Blachman.1967.p164,Ray.1973.p1469,Ray.1973.p1748,Beck.1984.p467,Sundholm.1991.p91,Joensson.1993.p4113,FEI.2003.p549,Chen.2012.p1}. Table~\ref{tab:HpC} gives the hyperfine interaction constants of these states for Be I from the MCDHF/RCI method to compare with data from other calculations and experiments. As can be seen from Table~\ref{tab:HpC}, the present MCDHF/RCI results agree well with the experimental values and the differences are all less than 0.1\%, which is similar to other theoretical values. The $rhfs$ code that we used to calculate the hyperfine structure does not include the QED correction, which is an important part of the difference. Meanwhile, the accuracy of the computed $B$ is limited by the accuracy of the corresponding $Q$ value. Here we used a new  value recommended by Puchalski $et~al.$~\cite{Puchalski.2021.p}. Additionally, some researchers have mentioned the contribution of high order QED effects~\citep{Sun.2023.p}, the Bohr-Weisskopf effect~\citep{Karpeshin.2015.p66} and other  contributions to the hyperfine structure parameter, which could all explain the remaining differences between our results and experiments.

\subsection{Isotope shifts}\label{subsec:isos}

Mass shift parameters and field shift parameters  are listed in Table~\ref{tab:HpI} for the lowest 99 levels of Be~I. The level mass shift and field shift
can be evaluated for any isotope of beryllium through Eqs.(8) and (9). The theoretical transition isotope shifts can be estimated from the corresponding level isotope shift differences. The nuclear masses are respectively, 11.02166108(26)~u for $^{11}\rm{Be}$, 10.01353469(9)~u for $^{10}\rm{Be}$ and 9.01218306(8)~ u for $^{9}\rm{Be}$~\citep{Wang.2021.p30003}. Let us consider the transition $2s\,3d~^{1}D_2-~2s^{2}~^{1}S_{0}$ of the $^{10}\rm{Be}-~^{9}{\rm{Be}}$ isotope pair as an example. Combining Eq.(8) with the calculated mass shift parameter $K_{\rm{NMS}} = 51~842~\rm{GHz~u}$ and $K_{\rm{SMS}} = 1~478.3~\rm{GHz~u}$ for $2s\,3d~^{1}D_2$ and $K_{\rm{NMS}} = 52~902~\rm{GHz~u}$ and $K_{\rm{SMS}} = 1~684.3~\rm{GHz~u}$ for $2s^{2}~^{1}S_{0}$, we deduce the normal mass shift $\delta \nu_{\rm{NMS}} = 11.757~\rm{GHz}$ and the specific mass shift $\delta \nu_{\rm{SMS}} = 2.286~\rm{GHz}$ for this transition. In comparison with the mass shift, the field shift is usually small for light atoms. The field shift is indeed estimated to be $\delta \nu_{\rm{FS}} = -0.006~\rm{GHz}$ for this transition, using $\delta \left \langle r^2 \right \rangle$ value provided by~\citet{Angeli.2013.p69}. The field shifts are rather small and can be neglected in our isotope shift calculations. To assess the uncertainties of the calculated isotope shift parameters, we compare the results from $\text{AS}_{14}$ and $\text{AS}_{15}$ based on $\text{MR}_7$. The normal and field mass shift parameters have a mean numerical uncertainty of 0.01$\%$, while that for the specific mass shift is somewhat larger at 0.1$\%$, due to its higher sensitivity to electron correlation effects.

\begin{figure}
    \centering
    \includegraphics[width=8.5cm]{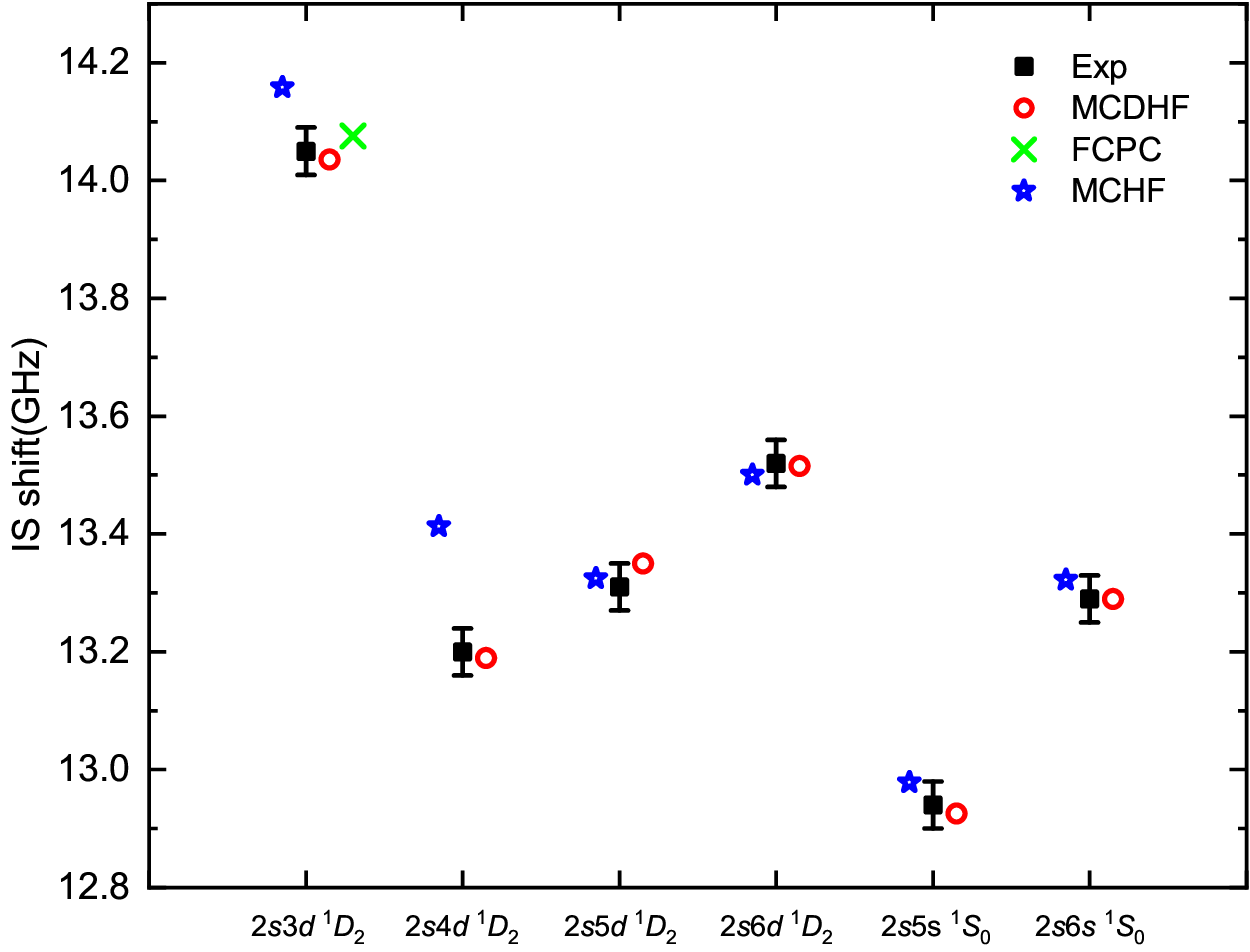}
    \caption{Comparison of transition isotope shifts of the $^{10}\rm{Be}-~^{9}{\rm{Be}}$ from the present MCDHF/RCI calculation with those from experimental measurement and MCHF method~\citep{Wen.1988.p4207}, FCPC method~\citep{Chung.1993.p1944}. In each case, the initial state is the $2s^{2}~^{1}S_{0}$ ground state.} 
    \label{fig:ISexp}
\end{figure}

Puchalski $et~al.$~\cite{Puchalski.2014.p} calculated the isotope mass shift of the $^{11}\rm{Be}-~^{9}{\rm{Be}}$
isotope pair for the $2s\,3s~^{1}S_0-~2s^{2}~^{1}S_{0}$ and $2s\,2p~^{1}P_1^{\circ}-~2s^{2}~^{1}S_{0}$ transitions through the ECG method. Their results for these two transitions are 18.90625(4)~GHz and 16.01529(3)~GHz, to be compared with our MCDHF/RCI values of 18.882~GHz and 16.007~GHz, respectively. A good agreement is achieved and the deviations are less than 0.2\%. Furthermore, ~\citet{Wen.1988.p4207} measured some isotope mass shifts of the $^{10}\rm{Be}-~^{9}{\rm{Be}}$ isotope pair for several transitions from $2s^{2}~^{1}S_{0}$ to $2s\,ns~^{1}S_0$ or $2s\,nd~^{1}D_2$ using high-resolution multiphoton-resonance-ionization mass spectroscopy. The experimental uncertainty is 40 MHz for the isotope shifts of all these transitions. Isotope shifts from the present MCDHF/RCI calculation are plotted in Fig.~\ref{fig:ISexp} along with experimental values and other theoretical values~\citep{Chung.1993.p1944}.
As can be seen from this figure, our MCDHF/RCI values agree well with the experimental ones. All our calculated isotope shifts fall within the experimental error bars. For most of the states, the transition isotope shifts calculated by~\citet{Wen.1988.p4207}
using the MCHF method are in good agreement with their measured values. 
However, theory-observation deviations of their work, beyond the range of the experimental error bar, appear for $2s^{2}~^{1}S_{0}-~2s\,3d~^{1}D_2$ (0.1 GHz) and $2s^{2}~^{1}S_{0}-~2s\,4d~^{1}D_2$ (0.21 GHz). 
For $2s^{2}~^{1}S_{0}-~2s\,3d~^{1}D_2$, a good agreement is found between the value calculated by Chung and Zhu and the measured value provided by Wen et al. However, the calculations of Chung and Zhu only focused on $n \leq 3$ levels. 

\section{Summary} \label{sec:summary}

Excitation energies, transition rates, lifetimes, Land${\acute{\rm{e}}}$ $g$-factors, hyperfine interaction constants and isotope shifts are reported for the 99 lowest levels of neutral beryllium. 
The influence of both the orbital active space and the multireference on excitation energies and transition rates has been discussed in detail. In order to achieve spectroscopic accuracy, the expansion of active space and multireference plays a particularly significant role and needs to be investigated carefully. 
Triple and quadruple excitations introduced through the extension of the multireference expansion are shown to be crucial to improve the agreement between different gauges used for evaluating the transition data.
The high accuracy of the present MCDHF/RCI results
is confirmed through two uncertainty estimation methods and comparisons with the available experimental and other theoretical results.

For the energy levels, the total average difference with the standard deviation from the NIST values is 7.08 $\pm$ 1.14 $\rm{cm^{-1}}$ (0.011$\%$ $\pm$ 0.003$\%$), which is highly satisfactory. For transition rates, two uncertainty estimation methods are used to estimate the accuracy of the present MCDHF/RCI line strengths. Comparison with high accuracy oscillator strengths obtained using the ECG method has also been made. The relative differences between the MCDHF/RCI oscillator strengths and those of the ECG method are within 2\% for most of the data. In addition, radiative lifetimes have been compared with previous experimental and theoretical values and a good agreement is observed, especially for some recent  measurements. The differences between the hyperfine interaction constants obtained with the present MCDHF/RCI method and the experimental values are less than 0.1\%. Our calculated isotope shifts fall within the experimental error bars.

The present work has significantly increased the amount of accurate data for the beryllium atom. These comprehensive data, especially transition rates, lifetimes, hyperfine interaction constants and isotope shifts, extend and improve the accuracy of atomic data for Be I and can be reliably applied in line identification and diagnostics for various astrophysical and laboratory plasma. The present work represents the most comprehensive and large-scale calculation employing MCDHF/RCI method to date, and can also be considered as a benchmark for other calculations.

\begin{acknowledgments}
We acknowledge the support from the National Key Research and Development Project, China (Grant No. 2022YFA1602303, 2022YFA1602500), the National Natural Science Foundation of China (Grants No. 12074081, No. 12104095, No. 12393824). P. J. acknowledges support from the Swedish Research Council (VR 2023-05367). A. M. A. acknowledges support from the Swedish Research Council (VR 2020-03940) and from the Crafoord Foundation via the Royal Swedish Academy of Sciences (CR 2024-0015).
\end{acknowledgments}

\vspace{1em}
\noindent This is the accepted manuscript of an article published in the \textit{Journal of Physical and Chemical Reference Data}. The final authenticated version will be available online at: https://doi.org/10.1063/5.0285265.

\section*{Data Availability}
The data that supports the findings of this study are available within the article and its supplementary material.

\clearpage

\clearpage
\onecolumngrid

\normalsize
\begin{longtable}{ccccc@{\hspace{1em}}lcccc}
\caption{Excitation energies (in cm$^{-1}$) and radiative lifetimes (in s) for the 99 lowest levels of the $n \leq 7$ configurations in Be I. $E_{\rm{MCDHF}}$: the present MCDHF/RCI excitation energies; $E_{\rm{Adjusted}}$: excitation energies corrected according to experimental and theoretical values; $E_{\rm{NIST}}$: the compiled values from the NIST database~\citep{Kramida.2023.p}; $\Delta E_{\rm{MCDHF}}$: energy differences (in $\rm{cm^{-1}}$) between $E_{\rm{MCDHF}}$ and $E_{\rm{NIST}}$; $\tau_{\rm MCDHF}^l$: the present MCDHF/RCI lifetimes in length form; $u(\tau^l)/\tau^l$: the relative uncertainty (in \%) of the present MCDHF/RCI lifetimes in length form; $LS$ composition: the $LS$ eigenvector compositions.\label{tab:Level}} \\

\toprule
Key & Level & $J\pi$ & $E_{\rm MCDHF}$ & $E_{\rm Adjusted}$ & $E_{\rm NIST}$ & $\Delta E_{\rm MCDHF}$ & $\tau_{\rm MCDHF}^l$ & $u(\tau^l)/\tau^l$ & $LS$ Composition \\
\midrule
\endfirsthead

\multicolumn{10}{c}{{\bfseries \tablename\ \thetable{} -- continued from previous page}} \\
\toprule
Key & Level & $J\pi$ & $E_{\rm MCDHF}$ & $E_{\rm Adjusted}$ & $E_{\rm NIST}$ & $\Delta E_{\rm MCDHF}$ & $\tau_{\rm MCDHF}^l$ & $u(\tau^l)/\tau^l$ & $LS$ Composition \\
\midrule
\endhead

\midrule \multicolumn{10}{r}{{Continued on next page}} \\
\endfoot

\bottomrule
\endlastfoot
1   & $2s^{2}~^{1}S_{0}$           & 0+   &       0  &  \nodata   &  \nodata   & \nodata  &  \nodata              &  \nodata    & 0.87 + 0.09~$2p^{2}~^{1}S$            \\
2   & $2s\,2p~^{3}P_{0}^{\circ}$   & 0$-$ &  21~973  &  21~978.31 & 21~978.31   & -4.86   &  \nodata              &  \nodata    & 0.93                                  \\
3   & $2s\,2p~^{3}P_{1}^{\circ}$   & 1$-$ &  21~974  &  21~978.92 & 21~978.925  & -5.42   &  3.593$\times10^0$    &  136        & 0.93                                  \\
4   & $2s\,2p~^{3}P_{2}^{\circ}$   & 2$-$ &  21~976  &  21~981.26 & 21~981.26   & -5.18   &  3.592$\times10^3$    &  0.01       & 0.93                                  \\
5   & $2s\,2p~^{1}P_{1}^{\circ}$   & 1$-$ &  42~564  &  42~565.45 & 42~565.4502 & -1.19   &  1.805$\times10^{-9}$ &  0.48       & 0.73 + 0.12~$2s\,3p~^{1}P^{\circ}$    \\
6   & $2s\,3s~^{3}S_{1}$           & 1+   &  52~072  &  52~080.94 & 52~080.943  & -8.76   &  6.712$\times10^{-9}$ &  0.32       & 0.96                                  \\
7   & $2s\,3s~^{1}S_{0}$           & 0+   &  54~668  &  54~677.35 & 54~677.349  & -8.87   &  2.885$\times10^{-8}$ &  0.48       & 0.95                                  \\
8   & $2p^{2}~^{1}D_{2}$           & 2+   &  56~875  &  56~882.55 & 56~882.5474 & -7.12   &  1.636$\times10^{-4}$ &  2.53       & 0.52 + 0.27~$2s\,3d~^{1}D$            \\
\end{longtable}
\clearpage

\normalsize
\begin{longtable}{l@{\hspace{0.4em}}l@{\hspace{0.4em}}c@{\hspace{0.4em}}c@{\hspace{0.4em}}c@{\hspace{0.5em}}c@{\hspace{0.5em}}c@{\hspace{0.5em}}c@{\hspace{0.5em}}c@{\hspace{0.5em}}c@{\hspace{0.5em}}c@{\hspace{0.5em}}c@{\hspace{0.4em}}c@{\hspace{0.4em}}c}
\caption{Transition wavelengths $\lambda$ (vacuum, in~{\AA}), transition rates $A$ (in s$^{-1}$), weighted oscillator strengths $gf$ (dimensionless), line strengths $S$ (in a.u.) and two estimated uncertainties Acc among the lowest 99 levels listed in Table~\ref{tab:Level}. $\lambda$ is transition wavelength (in \AA). Type is the type of the multipole, and BF is the branching fraction from the upper level. $A^l$, $gf^l$ and $S^l$ are, respectively, transition rates, weighted oscillator strengths, and line strengths in the length ($l$) form from the present MCDHF/RCI calculations. $A^v$, $gf^v$ and $S^v$ are, respectively, transition rates, weighted oscillator strengths, and line strengths in the velocity ($v$) form from the present MCDHF/RCI calculations. CF is the cancellation factor. The penultimate column ($\rm{Acc_1}$) represents the estimated accuracies of the $S$ values using the uncertainty estimation method suggested by ~\citet{Kramida.2024.p}. The last column ($\rm{Acc_2}$) represents the estimated accuracies of the $S$ values using the quantitative and qualitative evaluation approach~\citep{Rynkun.2022.p82,Kitoviene.2024.p}. Only transitions with BF $\geq~10^{-5}$ are presented. 
(This table is available in its entirety in machine-readable form.)\label{tab:Tr}} \\

\toprule
$i$ & $j$ &  $\lambda$ & Type &  $\text{BF}$&  $A^l$ &  $gf^l$ &  $S^l$ &  $A^v$ & $gf^v$ &  $S^v$ & $\text{CF}$ & $\rm{Acc_1}$ &  $\rm{Acc_2}$ \\
\midrule
\endfirsthead

\multicolumn{14}{c}{{\bfseries \tablename\ \thetable{} -- continued from previous page}} \\
\toprule
$i$ & $j$ &  $\lambda$ & Type &  $\rm BF$&  $A^l$ &  $gf^l$ &  $S^l$ &  $A^v$ & $gf^v$ &  $S^v$ & $\text{CF}$ & $\rm{Acc_1}$ &  $\rm{Acc_2}$ \\
\midrule
\endhead

\midrule \multicolumn{14}{r}{{Continued on next page}} \\
\endfoot

\bottomrule
\endlastfoot
  1 &  3 & 4~549.81 & E1 & 1.000$\times10^{0}$  & 2.783$\times10^{-1}$ & 2.591$\times10^{-9}$  & 3.881$\times10^{-8}$ & 1.176$\times10^{-1}$ & 1.095$\times10^{-9}$ & 1.640$\times10^{-8}$ & 2.52$\times10^{-10}$ &  E  & E  \\
  1 &  4 & 4~549.33 & M2 & 1.000$\times10^{0}$  & 2.784$\times10^{-4}$ & 4.319$\times10^{-12}$ & 1.819$\times10^{2}$  & \nodata              & \nodata              & \nodata              & \nodata              &  AA & AA \\
  1 &  5 & 2~349.32 & E1 & 1.000$\times10^{0}$  & 5.539$\times10^{8}$  & 1.375$\times10^{0}$   & 1.063$\times10^{1}$  & 5.535$\times10^{8}$  & 1.374$\times10^{0}$  & 1.063$\times10^{1}$  & 4.06$\times10^{-1}$  &  AA & AA \\
  1 &  8 & 1~758.01 & E2 & 2.655$\times10^{-1}$ & 1.624$\times10^{3}$  & 3.761$\times10^{-6}$  & 1.217$\times10^{2}$  & 1.623$\times10^{3}$  & 3.759$\times10^{-6}$ & 1.216$\times10^{2}$  & 6.62$\times10^{-1}$  &  AA & AA \\
  1 & 15 & 1~661.48 & E1 & 4.906$\times10^{-1}$ & 7.007$\times10^{6}$  & 8.699$\times10^{-3}$  & 4.758$\times10^{-2}$ & 6.999$\times10^{6}$  & 8.689$\times10^{-3}$ & 4.753$\times10^{-2}$ & 1.19$\times10^{-2}$  &  AA & AA \\
  1 & 25 & 1~491.76 & E1 & 7.112$\times10^{-2}$ & 2.591$\times10^{5}$  & 2.594$\times10^{-4}$  & 1.274$\times10^{-3}$ & 2.592$\times10^{5}$  & 2.594$\times10^{-4}$ & 1.274$\times10^{-3}$ & 1.28$\times10^{-3}$  &  AA & AA \\
  1 & 39 & 1~426.12 & E1 & 2.542$\times10^{-1}$ & 9.266$\times10^{5}$  & 8.476$\times10^{-4}$  & 3.979$\times10^{-3}$ & 9.262$\times10^{5}$  & 8.472$\times10^{-4}$ & 3.978$\times10^{-3}$ & 1.14$\times10^{-2}$  &  AA & AA \\
  1 & 55 & 1~394.10 & E1 & 1.121$\times10^{-4}$ & 1.582$\times10^{2}$  & 1.383$\times10^{-7}$  & 6.346$\times10^{-7}$ & 1.583$\times10^{2}$  & 1.383$\times10^{-7}$ & 6.349$\times10^{-7}$ & 7.30$\times10^{-6}$  &  B+ & AA \\
\end{longtable}
\clearpage

\begin{table*}
\renewcommand{\arraystretch}{1.2}
\caption{Comparison between present calculations and available theoretical oscillator strengths for the beryllium atom. \label{tab:TrC}}
\begin{ruledtabular}
\begin{tabular}{llccccc}
Lower State & Upper state & $f^l_{\rm{MCDHF}}$ & $f^v_{\rm{MCDHF}}$ & $f_{\rm{ECG}}\footnote{$f_{\rm{ECG}}$: values from ECG method~\citep{Nasiri.2021.p}}$ & $f_{\rm{BCICP}}$\footnote{$f_{\rm{BCICP}}$: values from BCICP method~\citep{Chen.1998.p4523}} & $f_{\rm{MCHF}}$\footnote{$f_{\rm{MCHF}}$: values from MCHF method~\citep{FroeseFischer.2004.p1}} \\ \hline
$2s^{2}~^{1}S_{0}$	        & $2s\,2p~^{1}P_{1}^{\circ}$	&  1.375$\times10^{0}$	& 1.374$\times10^{0}$ 	& 1.374$\times10^{0}$	  & 1.375$\times10^{0}$ 	& 1.380$\times10^{0}$  \\
$2s^{2}~^{1}S_{0}$	        & $2s\,3p~^{1}P_{1}^{\circ}$	&  8.699$\times10^{-3}$	& 8.689$\times10^{-3}$	& 8.582$\times10^{-3}$	& 9.01$\times10^{-3}$ 	& 8.985$\times10^{-3}$ \\
$2s^{2}~^{1}S_{0}$	        & $2s\,4p~^{1}P_{1}^{\circ}$	&  2.594$\times10^{-4}$	& 2.594$\times10^{-4}$	& 2.707$\times10^{-4}$	& 2.30$\times10^{-4}$ 	&                      \\
$2s^{2}~^{1}S_{0}$	        & $2s\,5p~^{1}P_{1}^{\circ}$	&  8.476$\times10^{-4}$	& 8.472$\times10^{-4}$	& 8.600$\times10^{-4}$	& 8.10$\times10^{-4}$ 	&                      \\
$2s^{2}~^{1}S_{0}$	        & $2s\,6p~^{1}P_{1}^{\circ}$	&  7.822$\times10^{-4}$	& 7.818$\times10^{-4}$	& 7.911$\times10^{-4}$	& 7.50$\times10^{-4}$ 	&                      \\
$2s^{2}~^{1}S_{0}$	        & $2s\,7p~^{1}P_{1}^{\circ}$	&  5.995$\times10^{-4}$	& 5.995$\times10^{-4}$	& 6.108$\times10^{-4}$	& 5.90$\times10^{-4}$ 	&                      \\
$2s\,2p~^{1}P_{1}^{\circ}$	& $2s\,3s~^{1}S_{0}$	        &  1.181$\times10^{-1}$	& 1.179$\times10^{-1}$	& 1.179$\times10^{-1}$	& 1.18$\times10^{-1}$ 	& 1.147$\times10^{-1}$ \\
\end{tabular}
\end{ruledtabular}
\end{table*}
\clearpage

\begin{table*}
\renewcommand{\arraystretch}{1.3}
\caption{Calculated lifetimes of Be I levels and comparison with previous results. The experimental lifetimes are highlighted in boldface when the present MCDHF/RCI values fall within the experimental error.\label{tab:Lif}}
\begin{ruledtabular}
\begin{tabular}{ccccll}
\multicolumn{2}{l}{Upper level} & \multicolumn{2}{l}{Lifetime(ns)} \\
\cline{1-2} \cline{3-6}
Config. & Term & \multicolumn{2}{l}{This work} & \multicolumn{2}{l}{Previous} \\
\cline{3-4} \cline{5-6} 
 &  & $\tau_{\rm MCDHF}^l$ & $\tau_{\rm MCDHF}^v$ & Exp.$^{*}$ & Cal.$^{\dagger}$ \\ \hline
$2s\,2p$ & $^{1}P^{\circ}$ &  1.81  &  1.81 &  $2.05(6)^{\text{a}}$,$2.3(1)^{\text{c}}$,$2.5(2)^{\text{d}}$,$\textbf{1.80(15)}^{\text{f}}$,$\textbf{1.77(5)}^{\text{i}}$,$\textbf{1.85(4)}^{\text{j}}$  &  $1.79^{\text{l}}$,$1.73^{\text{m}}$,$1.79^{\text{n}}$,$1.7^{\text{k}}$   \\ 
$2s\,3s$ & $^{3}S$         &  6.71  &  6.71 &  $5.95(16)^{\text{a}}$,$7.3(2)^{\text{c}}$,$6.2(2)^{\text{d}}$,$6.4(2)^{\text{h}}$                                                                        &  $6.70^{\text{l}}$,$6.65^{\text{m}}$,$6.61^{\text{n}}$,$6.7^{\text{k}}$   \\ 
$2p^{2}$ & $^{3}P$         &  2.36  &  2.36 &  $\textbf{2.42(6)}^{\text{a}}$,$\textbf{2.3(2)}^{\text{c}}$,$\textbf{2.4(1)}^{\text{d}}$                                                                  &  $2.31^{\text{l}}$,$2.37^{\text{n}}$,$2.2^{\text{k}}$                     \\ 
$2s\,3d$ & $^{3}D$         &  5.23  &  5.23 &  $5.00(14)^{\text{a}}$,$4.9(2)^{\text{c}}$,$\textbf{5.4(2)}^{\text{d}}$,$\textbf{5.2(2)}^{\text{h}}$                                                      &  $5.18^{\text{l}}$,$5.26^{\text{m}}$,$5.14^{\text{n}}$,$5.6^{\text{k}}$   \\ 
$2s\,3d$ & $^{1}D$         & 12.00  & 12.01 &  $\textbf{12.0(3)}^{\text{a}}$,$10.0(2)^{\text{c}}$,$\textbf{12.3(4)}^{\text{d}}$,$10.4(10)^{\text{e}}$                                                   &  $12.3^{\text{l}}$,$12.2^{\text{m}}$,$12.5^{\text{n}}$,$10.6^{\text{k}}$  \\
\end{tabular}
\end{ruledtabular}

\vspace{1mm}
\begin{flushleft}
$^{*}$~The experimental lifetimes from $^\text{a}$~\citet{Bergstroem.1969.p721}, $^\text{b}$~\citet{Bromander.1969.p523}, $^\text{c}$~\citet{Andersen.1969.p76}, $^\text{d}$~\citet{Bromander.1971.p61}, $^\text{e}$~\citet{Andersen.1971.p52}, $^\text{f}$~\citet{Hontzeas.1972.p55}, $^\text{g}$~\citet{Poulsen.1975.p1393}, $^\text{h}$~\citet{Kerkhoff.1980.p11}, $^\text{i}$~\citet{Irving.1999.p137}, $^\text{j}$~\citet{Schnabel.2000.p}, $^\text{k}$~\citet{Wang.2018.p112}.\\
$^{\dagger}$~The calculated lifetimes from $^\text{l}$~\citet{Moccia.1985.p3537}, $^\text{m}$~\citet{Chang.1990.p4922}, $^\text{n}$~\citet{Fischer.1997.p3333}, $^\text{k}$~\citet{Wang.2018.p112}.
\end{flushleft}
\end{table*}
\clearpage

\normalsize
\begin{longtable}{c@{\hspace{1em}}c@{\hspace{1.5em}}c@{\hspace{1.5em}}c@{\hspace{1.5em}}c@{\hspace{1.5em}}c@{\hspace{1.5em}}c@{\hspace{1.5em}}c@{\hspace{1.5em}}r@{\hspace{1.5em}}c@{\hspace{1.5em}}c}
\caption{Land$\acute{\rm{e}}$ $g$-factors $g_J$ and $LS$-coupling values $g_J(LS)$, hyperfine interaction constants $A_J$ and $B_J$ (both in MHz), mass shift parameters $K_{\rm NMS}$ and $K_{\rm SMS}$ (in GHz u), field shift parameters $F^{\rm (0)ved}$ (in GHz/$\rm{fm^2}$) and $F^{\rm (1)ved}$ (in GHz/$\rm{fm^4}$) among the lowest 99 levels listed in Table~\ref{tab:Level}.\label{tab:HpI}} \\

\toprule
Key & Level& $J\pi$ & $g_J$ & $g_J(LS)$ & $A_J$ & $B_J$ & $K_{\rm NMS}$ & $K_{\rm SMS}$ &  $F^{\rm (0)ved}$ & $F^{\rm (1)ved}$ \\
\midrule
\endfirsthead

\multicolumn{11}{c}{{\bfseries \tablename\ \thetable{} -- continued from previous page}} \\
\toprule
Key & Level& $J\pi$ & $g_J$ & $g_J(LS)$ & $A_J$ & $B_J$ & $K_{\rm NMS}$ & $K_{\rm SMS}$ &  $F^{\rm (0)ved}$ & $F^{\rm (1)ved}$ \\
\midrule
\endhead

\midrule \multicolumn{11}{r}{{Continued on next page}} \\
\endfoot

\bottomrule
\endlastfoot
1   & $2s^{2}~^{1}S_{0}$           & 0+   &  \nodata & \nodata  &  \nodata &   \nodata & 52~902 & 	1~684.3	 & 0.70052	 & -1.4382$\times10^{-5}$  \\
2   & $2s\,2p~^{3}P_{0}^{\circ}$   & 0$-$ &  \nodata & \nodata  &  \nodata &   \nodata & 52~540 & 	965.23   & 0.69066	 & -1.4179$\times10^{-5}$  \\
3   & $2s\,2p~^{3}P_{1}^{\circ}$   & 1$-$ &	  1.5000 &	1.5000  &  -139.27 &	 -0.7260 & 52~541 & 	965.27   & 0.69066	 & -1.4179$\times10^{-5}$  \\
4   & $2s\,2p~^{3}P_{2}^{\circ}$   & 2$-$ &	  1.5000 &	1.5000  &  -124.47 &	  1.4521 & 52~541 & 	965.10	 & 0.69066	 & -1.4179$\times10^{-5}$  \\
5   & $2s\,2p~^{1}P_{1}^{\circ}$   & 1$-$ &	  1.0000 &	1.0000  &  -13.836 &	  0.8452 & 52~202 & 	1~592.7	 & 0.69119	 & -1.4190$\times10^{-5}$  \\
6   & $2s\,3s~^{3}S_{1}$           & 1+   &	  2.0000 &	2.0000  &  -332.25 &	 	0.0000 & 52~046 & 	1~694.9	 & 0.69616	 & -1.4292$\times10^{-5}$  \\
7   & $2s\,3s~^{1}S_{0}$           & 0+   &  \nodata & \nodata  &  \nodata &   \nodata & 52~003 & 	1~649.4	 & 0.69574	 & -1.4284$\times10^{-5}$  \\
8   & $2p^{2}~^{1}D_{2}$           & 2+   &	  1.0000 &	1.0000  &  -13.938 &	  1.7418 & 51~967 & 	540.79   & 0.68434	 & -1.4049$\times10^{-5}$  \\
9   & $2s\,3p~^{3}P_{0}^{\circ}$   & 0$-$ &  \nodata &  0.0000  &  \nodata &   \nodata & 51~933 & 	1~576.5	 & 0.69464	 & -1.4261$\times10^{-5}$  \\
10  & $2s\,3p~^{3}P_{1}^{\circ}$   & 1$-$ &	  1.5000 &	1.5000  &  -154.22 &	 -0.1048 & 51~933 & 	1~576.8  & 0.69464	 & -1.4261$\times10^{-5}$  \\
11  & $2s\,3p~^{3}P_{2}^{\circ}$   & 2$-$ &	  1.5000 &	1.5000  &  -152.07 &	  0.2098 & 51~933 & 	1~576.7	 & 0.69464	 & -1.4261$\times10^{-5}$  \\
12  & $2p^{2}~^{3}P_{0}$           & 0+   &  \nodata & \nodata  &  \nodata &   \nodata & 51~920 & 	545.61   & 0.67749	 & -1.3909$\times10^{-5}$  \\
13  & $2p^{2}~^{3}P_{1}$           & 1+   &	  1.5000 &	1.5000  &   18.101 &	  0.7337 & 51~920 & 	545.29   & 0.67749	 & -1.3909$\times10^{-5}$  \\
14  & $2p^{2}~^{3}P_{2}$           & 2+   &	  1.5000 &	1.5000  &   3.3859 &	 -1.4680 & 51~920 & 	545.01   & 0.67749	 & -1.3909$\times10^{-5}$  \\
15  & $2s\,3p~^{1}P_{1}^{\circ}$   & 1$-$ &	  1.0000 &	1.0000  &  -6.7039 &	  0.4272 & 51~912 & 	1~468.9	 & 0.69315	 & -1.4230$\times10^{-5}$  \\
16  & $2s\,3d~^{3}D_{1}$           & 1+   &	  0.5000 &	0.5000  &   150.60 &	  0.0159 & 51~881 & 	1~664.6	 & 0.69484	 & -1.4265$\times10^{-5}$  \\
17  & $2s\,3d~^{3}D_{2}$           & 2+   &	  1.1667 &	1.1667  &  -50.946 &	  0.0226 & 51~881 & 	1~664.6	 & 0.69484	 & -1.4265$\times10^{-5}$  \\
18  & $2s\,3d~^{3}D_{3}$           & 3+   &   1.3333 &  1.3333  &  -101.22 &    0.0456 & 51~881 & 	1~664.6	 & 0.69484	 & -1.4265$\times10^{-5}$  \\
19  & $2s\,3d~^{1}D_{2}$           & 2+   &	  1.0000 &	1.0000  &  -3.9842 &	  0.5005 & 51~842 & 	1~478.3	 & 0.69230	 & -1.4213$\times10^{-5}$  \\
20  & $2s\,4s~^{3}S_{1}$           & 1+   &   2.0000 &  2.0000  &  -318.76 &    0.0000 & 51~841 & 	1~673.2	 & 0.69557	 & -1.4280$\times10^{-5}$  \\
21  & $2s\,4s~^{1}S_{0}$           & 0+   &  \nodata & \nodata  &  \nodata &   \nodata & 51~830 & 	1~657.1	 & 0.69541	 & -1.4277$\times10^{-5}$  \\
22  & $2s\,4p~^{3}P_{0}^{\circ}$   & 0$-$ &  \nodata & \nodata  &  \nodata &   \nodata & 51~803 & 	1~628.0	 & 0.69505	 & -1.4269$\times10^{-5}$  \\
23  & $2s\,4p~^{3}P_{1}^{\circ}$   & 1$-$ &	  1.5000 &	1.5000  &  -155.26 &	 -0.0385 & 51~803 & 	1~628.2	 & 0.69505	 & -1.4269$\times10^{-5}$  \\
24  & $2s\,4p~^{3}P_{2}^{\circ}$   & 2$-$ &	  1.5000 &	1.5000  &  -154.71 &	  0.0772 & 51~803 & 	1~628.1	 & 0.69505	 & -1.4269$\times10^{-5}$  \\
25  & $2s\,4p~^{1}P_{1}^{\circ}$   & 1$-$ &	  1.0000 &	1.0000  &  -3.0264 &	  0.1808 & 51~799 & 	1~578.2	 & 0.69439	 & -1.4256$\times10^{-5}$  \\
26  & $2s\,4d~^{3}D_{1}$           & 1+   &	  0.5000 &	0.5000  &   154.01 &	  0.0063 & 51~784 & 	1~664.8	 & 0.69512	 & -1.4271$\times10^{-5}$  \\
27  & $2s\,4d~^{3}D_{2}$           & 2+   &	  1.1667 &	1.1667  &  -51.661 &	  0.0091 & 51~784 & 	1~664.8	 & 0.69511	 & -1.4271$\times10^{-5}$  \\
28  & $2s\,4d~^{3}D_{3}$           & 3+   &	  1.3333 &	1.3333  &  -103.01 &	  0.0183 & 51~784 & 	1~664.8	 & 0.69511	 & -1.4271$\times10^{-5}$  \\
29  & $2s\,4f~^{3}F_{4}^{\circ}$   & 4$-$ &	  1.2500 &	1.2500  &  -77.734 &	  0.0070 & 51~779 & 	1~651.5	 & 0.69521	 & -1.4273$\times10^{-5}$  \\
30  & $2s\,4f~^{3}F_{2}^{\circ}$   & 2$-$ &	  0.6667 &	0.6667  &   103.52 &	  0.0048 & 51~779 & 	1~651.4	 & 0.69521	 & -1.4273$\times10^{-5}$  \\
31  & $2s\,4f~^{3}F_{3}^{\circ}$   & 3$-$ &	  1.0808 &	1.0833  &  -55.896 &	  0.0053 & 51~779 & 	1~651.5	 & 0.69521	 & -1.4273$\times10^{-5}$  \\
32  & $2s\,4f~^{1}F_{3}^{\circ}$   & 3$-$ &	  1.0025 &	1.0000  &   29.903 &	  0.0069 & 51~779 & 	1~651.9	 & 0.69521	 & -1.4273$\times10^{-5}$  \\
33  & $2s\,4d~^{1}D_{2}$           & 2+   &	  1.0000 &	1.0000  &  -1.1536 &	  0.1466 & 51~770 & 	1~626.5	 & 0.69441	 & -1.4256$\times10^{-5}$  \\
34  & $2s\,5s~^{3}S_{1}$           & 1+   &	  2.0000 &	2.0000  &  -315.10 &	  0.0000 & 51~767 & 	1~665.4	 & 0.69541	 & -1.4277$\times10^{-5}$  \\
35  & $2s\,5s~^{1}S_{0}$           & 0+   &  \nodata & \nodata  &  \nodata &   \nodata & 51~763 & 	1~658.1	 & 0.69533	 & -1.4275$\times10^{-5}$  \\
36  & $2s\,5p~^{3}P_{0}^{\circ}$   & 0$-$ &  \nodata & \nodata  &  \nodata &   \nodata & 51~749 & 	1~643.4	 & 0.69516	 & -1.4272$\times10^{-5}$  \\
37  & $2s\,5p~^{3}P_{1}^{\circ}$   & 1$-$ &   1.5000 &  1.5000  &  -154.34 &   -0.0185 & 51~749 & 	1~643.6	 & 0.69517	 & -1.4272$\times10^{-5}$  \\
38  & $2s\,5p~^{3}P_{2}^{\circ}$   & 2$-$ &	  1.5000 &	1.5000  &  -155.43 &	  0.0371 & 51~749 & 	1~643.6	 & 0.69517	 & -1.4272$\times10^{-5}$  \\
39  & $2s\,5p~^{1}P_{1}^{\circ}$   & 1$-$ &	  1.0000 &	1.0000  &  -2.8538 &	  0.0910 & 51~749 & 	1~619.0	 & 0.69483	 & -1.4265$\times10^{-5}$  \\
40  & $2s\,5d~^{3}D_{3}$           & 3+   &	  1.3333 &	1.3333  &  -103.54 &	  0.0091 & 51~740 & 	1~661.9	 & 0.69520	 & -1.4272$\times10^{-5}$  \\
41  & $2s\,5d~^{3}D_{2}$           & 2+   &	  1.1667 &	1.1667  &  -51.887 &	  0.0045 & 51~740 & 	1~661.9	 & 0.69520	 & -1.4272$\times10^{-5}$  \\
42  & $2s\,5d~^{3}D_{1}$           & 1+   &	  0.5000 &	0.5000  &   155.05 &	  0.0032 & 51~740 & 	1~661.9	 & 0.69520	 & -1.4272$\times10^{-5}$  \\
43  & $2s\,5f~^{3}F_{4}^{\circ}$   & 4$-$ &	  1.2500 &	1.2500  &  -77.843 &	  0.0042 & 51~738 & 	1~653.8	 & 0.69524	 & -1.4273$\times10^{-5}$  \\
44  & $2s\,5f~^{3}F_{2}^{\circ}$   & 2$-$ &	  0.6667 &	0.6667  &   103.72 &	  0.0029 & 51~738 & 	1~653.8	 & 0.69524	 & -1.4273$\times10^{-5}$  \\
45  & $2s\,5f~^{3}F_{3}^{\circ}$   & 3$-$ &	  1.0818 &	1.0833  &  -49.328 &	  0.0032 & 51~738 & 	1~653.8	 & 0.69524	 & -1.4273$\times10^{-5}$  \\
46  & $2s\,5f~^{1}F_{3}^{\circ}$   & 3$-$ &	  1.0015 &	1.0000  &   23.332 &	  0.0041 & 51~738 & 	1~654.3	 & 0.69524	 & -1.4273$\times10^{-5}$  \\
47  & $2s\,5g~^{3}G_{3}$           & 3+   &	  0.7500 &	0.7500  &   77.959 &	  0.0009 & 51~737 & 	1~656.7	 & 0.69527	 & -1.4274$\times10^{-5}$  \\
48  & $2s\,5g~^{3}G_{4}$           & 4+   &	  1.0278 &	1.0500  &  -77.979 &	  0.0011 & 51~737 & 	1~656.7	 & 0.69527	 & -1.4274$\times10^{-5}$  \\
49  & $2s\,5g~^{3}G_{5}$           & 5+   &	  1.2000 &	1.2000  &  -62.382 &	  0.0012 & 51~737 & 	1~656.7	 & 0.69527	 & -1.4274$\times10^{-5}$  \\
50  & $2s\,5g~^{1}G_{4}$           & 4+   &	  1.0222 &	1.0000  &   62.369 &	  0.0011 & 51~737 & 	1~656.7	 & 0.69527	 & -1.4274$\times10^{-5}$  \\
51  & $2s\,5d~^{1}D_{2}$           & 2+   &	  1.0000 &	1.0000  &  -0.4563 &	  0.0626 & 51~734 & 	1~648.7	 & 0.69490	 & -1.4266$\times10^{-5}$  \\
52  & $2s\,6s~^{3}S_{1}$           & 1+   &	  2.0000 &	2.0000  &  -313.71 &	  0.0000 & 51~732 & 	1~662.2	 & 0.69535	 & -1.4276$\times10^{-5}$  \\
53  & $2s\,6s~^{1}S_{0}$           & 0+   &  \nodata & \nodata  &  \nodata &   \nodata & 51~730 & 	1~658.1	 & 0.69530	 & -1.4275$\times10^{-5}$  \\
54  & $2s\,6p~^{3}P_{0}^{\circ}$   & 0$-$ &  \nodata & \nodata  &  \nodata &   \nodata & 51~722 & 	1~649.8	 & 0.69521	 & -1.4273$\times10^{-5}$  \\
55  & $2s\,6p~^{3}P_{1}^{\circ}$   & 1$-$ &	  1.4999 &	1.5000  &  -149.82 &	 -0.0102 & 51~722 & 	1~650.0	 & 0.69521	 & -1.4273$\times10^{-5}$  \\
56  & $2s\,6p~^{3}P_{2}^{\circ}$   & 2$-$ &	  1.5000 &	1.5000  &  -155.73 &	  0.0206 & 51~722 & 	1~649.9	 & 0.69521	 & -1.4273$\times10^{-5}$  \\
57  & $2s\,6p~^{1}P_{1}^{\circ}$   & 1$-$ &	  1.0001 &	1.0000  &  -6.8844 &	  0.0519 & 51~722 & 	1~636.2	 & 0.69502	 & -1.4269$\times10^{-5}$  \\
58  & $2s\,6d~^{3}D_{1}$           & 1+   &	  0.5000 &	0.5000  &   155.47 &	  0.0018 & 51~717 & 	1~660.2	 & 0.69523	 & -1.4273$\times10^{-5}$  \\
59  & $2s\,6d~^{3}D_{2}$           & 2+   &	  1.1667 &	1.1667  &  -52.005 &	  0.0026 & 51~717 & 	1~660.2	 & 0.69523	 & -1.4273$\times10^{-5}$  \\
60  & $2s\,6d~^{3}D_{3}$           & 3+   &	  1.3333 &	1.3333  &  -103.75 &	  0.0053 & 51~717 & 	1~660.2	 & 0.69523	 & -1.4273$\times10^{-5}$  \\
61  & $2s\,6f~^{3}F_{2}^{\circ}$   & 2$-$ &	  0.6667 &	0.6667  &   103.83 &	  0.0018 & 51~716 & 	1~655.1	 & 0.69525	 & -1.4274$\times10^{-5}$  \\
62  & $2s\,6f~^{3}F_{3}^{\circ}$   & 3$-$ &	  1.0816 &	1.0833  &  -51.330 &	  0.0020 & 51~716 & 	1~655.2	 & 0.69525	 & -1.4274$\times10^{-5}$  \\
63  & $2s\,6f~^{3}F_{4}^{\circ}$   & 4$-$ &	  1.2500 &	1.2500  &  -77.905 &	  0.0027 & 51~716 & 	1~655.2	 & 0.69525	 & -1.4274$\times10^{-5}$  \\
64  & $2s\,6f~^{1}F_{3}^{\circ}$   & 3$-$ &	  1.0018 &	1.0000  &   25.332 &	  0.0026 & 51~716 & 	1~655.5	 & 0.69525	 & -1.4274$\times10^{-5}$  \\
65  & $2s\,6g~^{3}G_{3}$           & 3+   &	  0.7500 &	0.7500  &   77.978 &	  0.0006 & 51~715 & 	1~657.0	 & 0.69527	 & -1.4274$\times10^{-5}$  \\
66  & $2s\,6g~^{3}G_{4}$           & 4+   &	  1.0279 &	1.0500  &  -77.989 &	  0.0007 & 51~715 & 	1~657.0	 & 0.69527	 & -1.4274$\times10^{-5}$  \\
67  & $2s\,6g~^{3}G_{5}$           & 5+   &	  1.2000 &	1.2000  &  -62.391 &	  0.0007 & 51~715 & 	1~657.0	 & 0.69527	 & -1.4274$\times10^{-5}$  \\
68  & $2s\,6g~^{1}G_{4}$           & 4+   &	  1.0221 &	1.0000  &   62.382 &	  0.0007 & 51~715 & 	1~657.0	 & 0.69527	 & -1.4274$\times10^{-5}$  \\
69  & $2s\,6h~^{3}H_{4}^{\circ}$   & 4$-$ &	  0.8000 &	0.8000  &   62.400 &	  0.0003 & 51~715 & 	1~657.2	 & 0.69527	 & -1.4274$\times10^{-5}$  \\
70  & $2s\,6h~^{3}H_{5}^{\circ}$   & 5$-$ &	  1.0181 &	1.0333  &  -62.405 &	  0.0003 & 51~715 & 	1~657.2	 & 0.69527	 & -1.4274$\times10^{-5}$  \\
71  & $2s\,6h~^{3}H_{6}^{\circ}$   & 6$-$ &	  1.1667 &	1.1667  &  -52.001 &	  0.0003 & 51~715 & 	1~657.2	 & 0.69527	 & -1.4274$\times10^{-5}$  \\
72  & $2s\,6h~^{1}H_{5}^{\circ}$   & 5$-$ &	  1.0152 &	1.0000  &   51.999 &	  0.0003 & 51~715 & 	1~657.2	 & 0.69527	 & -1.4274$\times10^{-5}$  \\
73  & $2s\,6d~^{1}D_{2}$           & 2+   &	  1.0000 &	1.0000  &  -0.1767 &	  0.0329 & 51~713 & 	1~654.2	 & 0.69508	 & -1.4270$\times10^{-5}$  \\
74  & $2s\,7s~^{3}S_{1}$           & 1+   &	  2.0000 &	2.0000  &  -313.08 &	  0.0000 & 51~712 & 	1~660.6	 & 0.69533	 & -1.4275$\times10^{-5}$  \\
75  & $2s\,7s~^{1}S_{0}$           & 0+   &  \nodata & \nodata  &  \nodata &   \nodata & 51~711 & 	1~658.0	 & 0.69529	 & -1.4274$\times10^{-5}$  \\
76  & $2s\,7p~^{3}P_{0}^{\circ}$   & 0$-$ &  \nodata & \nodata  &  \nodata &   \nodata & 51~705 & 	1~652.9	 & 0.69523	 & -1.4273$\times10^{-5}$  \\
77  & $2s\,7p~^{3}P_{2}^{\circ}$   & 2$-$ &	  1.5000 &	1.5000  &  -155.87 &	  0.0126 & 51~706 & 	1~653.0	 & 0.69524	 & -1.4273$\times10^{-5}$  \\
78  & $2s\,7p~^{3}P_{1}^{\circ}$   & 1$-$ &	  1.4982 &	1.5000  &  -128.84 &	 -0.0061 & 51~706 & 	1~653.0	 & 0.69524	 & -1.4273$\times10^{-5}$  \\
79  & $2s\,7p~^{1}P_{1}^{\circ}$   & 1$-$ &	  1.0018 &	1.0000  &  -27.611 &	  0.0321 & 51~706 & 	1~644.7	 & 0.69512	 & -1.4271$\times10^{-5}$  \\
80  & $2s\,7d~^{3}D_{3}$           & 3+   &	  1.3333 &	1.3333  &  -103.85 &	  0.0033 & 51~703 & 	1~659.3	 & 0.69525	 & -1.4273$\times10^{-5}$  \\
81  & $2s\,7d~^{3}D_{2}$           & 2+   &	  1.1667 &	1.1667  &  -52.048 &	  0.0016 & 51~703 & 	1~659.3	 & 0.69525	 & -1.4273$\times10^{-5}$  \\
82  & $2s\,7d~^{3}D_{1}$           & 1+   &	  0.5000 &	0.5000  &   155.68 &	  0.0012 & 51~703 & 	1~659.2	 & 0.69525	 & -1.4273$\times10^{-5}$  \\
83  & $2s\,7f~^{3}F_{4}^{\circ}$   & 4$-$ &	  1.2500 &	1.2500  &  -77.941 &	  0.0018 & 51~702 & 	1~656.0	 & 0.69526	 & -1.4274$\times10^{-5}$  \\
84  & $2s\,7f~^{3}F_{3}^{\circ}$   & 3$-$ &	  1.0802 &	1.0833  &  -59.309 &	  0.0013 & 51~702 & 	1~656.0	 & 0.69526	 & -1.4274$\times10^{-5}$  \\
85  & $2s\,7f~^{3}F_{2}^{\circ}$   & 2$-$ &	  0.6667 &	0.6667  &   103.89 &	  0.0012 & 51~702 & 	1~655.9	 & 0.69526	 & -1.4274$\times10^{-5}$  \\
86  & $2s\,7f~^{1}F_{3}^{\circ}$   & 3$-$ &	  1.0031 &	1.0000  &   33.309 &	  0.0017 & 51~702 & 	1~656.2	 & 0.69526	 & -1.4274$\times10^{-5}$  \\
87  & $2s\,7g~^{3}G_{3}$           & 3+   &	  0.7500 &	0.7500  &   77.989 &	  0.0004 & 51~701 & 	1~657.2	 & 0.69527	 & -1.4274$\times10^{-5}$  \\
88  & $2s\,7g~^{3}G_{4}$           & 4+   &	  1.0279 &	1.0500  &  -77.996 &	  0.0004 & 51~701 & 	1~657.2	 & 0.69527	 & -1.4274$\times10^{-5}$  \\
89  & $2s\,7g~^{3}G_{5}$           & 5+   &	  1.2000 &	1.2000  &  -62.397 &	  0.0005 & 51~701 & 	1~657.2	 & 0.69527	 & -1.4274$\times10^{-5}$  \\
90  & $2s\,7g~^{1}G_{4}$           & 4+   &	  1.0221 &	1.0000  &   62.391 &	  0.0004 & 51~701 & 	1~657.2	 & 0.69527	 & -1.4274$\times10^{-5}$  \\
91  & $2s\,7h~^{3}H_{4}^{\circ}$   & 4$-$ &	  0.8000 &	0.8000  &   62.403 &	  0.0002 & 51~702 & 	1~657.2	 & 0.69527	 & -1.4274$\times10^{-5}$  \\
92  & $2s\,7h~^{3}H_{5}^{\circ}$   & 5$-$ &	  1.0181 &	1.0333  &  -62.406 &	  0.0002 & 51~702 & 	1~657.2	 & 0.69527	 & -1.4274$\times10^{-5}$  \\
93  & $2s\,7h~^{3}H_{6}^{\circ}$   & 6$-$ &	  1.1667 &	1.1667  &  -52.002 &	  0.0002 & 51~701 & 	1~657.3	 & 0.69527	 & -1.4274$\times10^{-5}$  \\
94  & $2s\,7h~^{1}H_{5}^{\circ}$   & 5$-$ &	  1.0152 &	1.0000  &   52.001 &	  0.0002 & 51~702 & 	1~657.2	 & 0.69527	 & -1.4274$\times10^{-5}$  \\
95  & $2s\,7i~^{3}I_{5}$           & 5+   &	  0.8333 &	0.8333  &   52.004 &	  0.0001 & 51~702 & 	1~657.3	 & 0.69527	 & -1.4274$\times10^{-5}$  \\
96  & $2s\,7i~^{3}I_{6}$           & 6+   &	  1.0128 &	1.0238  &  -52.003 &	  0.0001 & 51~702 & 	1~657.4	 & 0.69527	 & -1.4274$\times10^{-5}$  \\
97  & $2s\,7i~^{3}I_{7}$           & 7+   &	  1.1429 &	1.1429  &  -44.572 &	  0.0001 & 51~701 & 	1~657.3	 & 0.69527	 & -1.4274$\times10^{-5}$  \\
98  & $2s\,7i~^{1}I_{6}$           & 6+   &	  1.0110 &	1.0000  &   44.573 &	  0.0001 & 51~702 & 	1~657.3	 & 0.69527	 & -1.4274$\times10^{-5}$  \\
99  & $2s\,7d~^{1}D_{2}$           & 2+   &	  1.0000 &	1.0000  &  -0.0632 &	  0.0196 & 51~701 & 	1~656.0	 & 0.69516	 & -1.4272$\times10^{-5}$  \\
\end{longtable}
\clearpage

\begin{table*}
\renewcommand{\arraystretch}{1.2}
\caption{Comparison of the hyperfine interaction constants (in MHz) of the $1s^2\,2s\,2p~^{3}P_{1,2}^{\circ}$ states for the $^{9}\text{Be}$ isotope.\label{tab:HpC}}
\begin{ruledtabular}
\begin{tabular}{lllll}
Method & $A_2$ & $A_1$ & $B_2$ & $B_1$ \\ \hline
$\rm{LC~MBPT}$\footnote{Linked-cluster many-body perturbation theory from~\citet{Ray.1973.p1469}}    &  -124.47   &  -139.10  &          &           \\
$\rm{HF + SDCI}$\footnote{Hartree-Fock and CI allowing all SD excitations to correlation orbitals of Slater type from~\citet{Beck.1984.p467}}  &  -124.76   &  -139.77  &          &           \\
$\rm{FE~MCHF}$\footnote{Finite-element multiconfiguration Hartree-Fock from~\citet{Sundholm.1991.p91}}   &  -124.50   &  -139.27  &          &           \\
$\rm{MCHF}$\footnote{MCHF from~\citet{Joensson.1993.p4113}}       &  -124.50   &  -139.35  &          &           \\
$\rm{RRV + CI}$\footnote{The Rayleigh-Ritz variational method combined with multi-CI from~\citet{FEI.2003.p549}}    &  -124.349  &  -139.012 &  1.44103 &  -0.72052 \\
$\rm{FCPC}$\footnote{FCPC method from~\citet{Chen.2012.p1}}\       &  -124.51   &  -139.36  &          &           \\
This work             &  -124.47   &  -139.27  &  1.4521    &  -0.7260    \\
$\rm{Experiment}$\footnote{The experimental hyperfine interaction constants from~\citet{Blachman.1967.p164}}    &  -124.5368(17) &  -139.373(12) &  1.429(8)   &     -0.753(44)      \\
\end{tabular}
\end{ruledtabular}
\end{table*}
\clearpage

\twocolumngrid

\bibliography{ref.bib}

\end{document}